\documentclass[a4paper,11pt]{article}
\usepackage{jheppub} 
\makeatletter
\gdef\@fpheader{}
\makeatother

\usepackage{amsfonts}
\usepackage{bbold}
\usepackage{booktabs}
\usepackage{xcolor}

\usepackage{tikz}
\usetikzlibrary{arrows.meta}
\definecolor{vpurple}{RGB}{125,40,145}
\definecolor{vorange}{RGB}{230,120,40}

\newcommand{\Etmiss}{E_T^{\text{miss}}}
\newcommand{\mttvis}{M_{t\bar{t}}^{\text{vis}}}
\newcommand{\ttbar}{t\bar{t}}
\newcommand{\mttwo}{m_{T2}}

\newcommand{\bsc}{Barcelona Supercomputing Center (BSC), Pla\c{c}a Eusebi G\"uell 1-3, 08034 Barcelona.}
\newcommand{\ifae}{Institut de F\'{\i}sica d'Altes Energies (IFAE), The Barcelona Institute of Science and Technology (BIST), Campus UAB, 08193 Bellaterra, Barcelona, Spain}
\newcommand{\icrea}{Institució Catalana de Recerca i Estudis Avançats (ICREA), Passeig Lluís Companys 23, 08010 Barcelona, Spain}
\newcommand{\ub}{Departament de Física Quàntica i Astrofísica, Universitat de Barcelona, C/ Martí i Franquès, 1-11, 08028 Barcelona}

\title{Quantum probes of  invisible particles in top-quark pair production at the LHC}

\author[1,2]{Blanca Belmonte,}
\author[3,4]{Diego Blas,}
\author[1]{Alba Cervera-Lierta,}
\author[3]{Francesco Montagno} 

\affiliation[1]{\bsc}
\affiliation[2]{\ub}
\affiliation[3]{\ifae}
\affiliation[4]{\icrea}

\emailAdd{fmontagno@ifae.es} 

\abstract{
Top-quark pair production at the LHC provides a bipartite spin system whose correlations are imprinted in the angular distributions of the decay products and can be studied with quantum-information tools. We investigate whether this spin structure can be used to probe invisible new physics in \(pp\to t\bar t+X\), where \(X\) is a spin-0 or spin-1 mediator motivated by top-philic simplified dark-matter models. Focusing on the dileptonic channel at the High-Luminosity LHC, we compare the sensitivity of the inclusive event yield, the visible invariant-mass distribution, an event-level estimator of the entanglement marker \(D\), and the fiducial spin correlation \(\mathcal D\) measured as a function of the visible invariant mass. We find that observables retaining spin information improve the expected exclusion sensitivity with respect to rate and kinematic information alone, with \(\mathcal D(M_{t\bar t}^{\rm vis})\) providing the strongest reach for both mediator spins. The same observables also enhance the discrimination between scalar and pseudoscalar couplings of the spin-zero mediator, with \(\mathcal D(M_{t\bar t}^{\rm vis})\) again providing the strongest sensitivity, particularly at larger mediator masses. These results show that quantum-information-motivated observables of the \(t\bar t\) system can be efficient tools to probe both the presence of invisible new physics and the structure of its couplings to top quarks.
}

\begin{document}
    \maketitle
\flushbottom

\section{Introduction} \label{sec:intro}

The top quark is the heaviest known elementary particle and the only quark that decays before hadronising, $\tau_t \ll \tau_{\rm QCD}$. Its spin information is therefore not washed out by non-perturbative dynamics, but is transferred directly to the angular distributions of its decay products. Top-quark pair production at the LHC consequently
offers something that no other collider process offers at comparable rates: a bipartite spin system whose quantum state can be reconstructed as a two-qubit density matrix $\rho_{t\bar t}$ and analysed with the tools of quantum information (QI), rather than
only through cross sections and kinematic distributions. Following the original proposals to measure spin entanglement in $t\bar t$
production~\cite{Afik2021TopEntanglement,Fabbrichesi2021Testing,Severi2022Quantum,AguilarSaavedra2022Improved}, entanglement between the top and antitop spins has now been observed by ATLAS~\cite{ATLAS:2023fsd} and CMS~\cite{CMSCollaboration2024,CMS:2024zkc}, making the LHC the
highest-energy system in which quantum correlations have been directly probed~\cite{Barr2024Quantum,Afik2025Quantum}.
 
Beyond testing quantum mechanics at high energies, QI observables are also sensitive probes of physics beyond the Standard Model (BSM). Since they are constructed from the full spin-correlation matrix, they are sensitive to the helicity structure of the production amplitude, which a new interaction can distort even in regions of phase space where the total rate remains close to its Standard Model (SM) value. This complementarity has been exploited to constrain SM Effective Field Theory (SMEFT) operators in $t\bar t$ production~\cite{Aoude2022Quantum,Severi2023Quantum,Fabbrichesi:2022ovb,Aoude:2025jzc} and in diboson final states~\cite{Aoude2023Probing}, to search for intermediate resonances with quantum observables~\cite{Maltoni2024Quantum}, and in prospective studies at future colliders~\cite{Maltoni2024QuantumTops,Goncalves2026Tripartite}.

More generally, spin-correlation observables have long been exploited in BSM searches, including experimental searches for heavy scalar and pseudoscalar states decaying to $t\bar t$~\cite{CMS:2025dzq}. In all of these cases, however, the new state is either purely virtual or, if produced on-shell, visible in the detector. 
Spin correlations have also been exploited in searches involving invisible particles. 
An early example is provided by light stops decaying to a nearly massless neutralino~\cite{Han2012,ATLAS2015}, where the invisible particles arise from the decay of a stop pair that mimics $t\bar t$ and the effect is captured through a single angular variable.
 
In this paper we address a different situation: a new particle that is produced on-shell but escapes the detector. We study $pp \to t\bar t + X$ at the LHC, where $X$ is an invisible spin-0 ($Y_0$) or spin-1 ($Y_1$) mediator, and ask whether a QI observable built from $\rho_{t\bar t}$ is sensitive to its presence. This final state is well motivated in simplified dark matter (DM) models, in which a top-philic mediator couples the SM to a dark sector and decays into a pair of DM
particles~\cite{Haisch2015Siplified,Haisch:2016gry,Arina2016Comprehensive,Backovic:2015soa}. 
Searches in the $t\bar t + E_T^{\rm miss}$ channel have been performed by ATLAS~\cite{Aad2023Constraints} and CMS~\cite{CMS2025Search}, and dedicated optimal observables sensitive to the CP nature of the mediator have been proposed at the phenomenological level~\cite{Buckley:2015ctj,Haisch:2016gry}. More recently, it has been shown that the dileptonic $t\bar t$ system can be reconstructed through a kinematic fit even in the presence of an invisible spin-0~\cite{Azevedo:2023xuc} or spin-1~\cite{Capucha:2026oqu} mediator, and that the resulting angular observables retain sensitivity to the presence of the mediator and to the spin and CP structure of its couplings.

The use of $t\bar t$ spin correlations to probe invisible new physics is not itself new, and has already been exploited experimentally in searches for dark matter produced in association with top quarks~\cite{CMS2025Search}. What has not, to the best of our knowledge, been explored in this context is an explicit quantum-information formulation of the $t\bar t$ subsystem in terms of its reduced spin density matrix, together with a systematic assessment of QI-motivated observables relative to rate and kinematic information. In addition, to the best of our knowledge, the case of an invisible spin-1 mediator has not been investigated experimentally in this context and provides a further novel aspect of the present study.

Among the many observables one could in principle construct to test for
BSM effects, the QI-inspired observable used in this work is singled out
by its basis independence. The scalar quantity
$D$ is invariant under rotations of the
reference-frame axes, so that its value does not depend on the convention
used to define the helicity frame and can be compared directly across
different analyses. This invariance holds for the density matrix of the full production process; once a fiducial region is defined by selection cuts, the extracted coefficients are acceptance-weighted averages and no longer admit a strict density-matrix interpretation. However, even in that situation, this observable is not an arbitrary function of the data: it is physics motivated and may encode information about the entanglement between the tops.
 
There is also a clear physical picture of why an invisible extra particle should show up in $\rho_{t\bar t}$. The observable considered here is built from a reduced, and therefore generally mixed, density matrix, and the mixedness has two distinct origins. The first is classical: the observed sample is an incoherent superposition of different production configurations, summed over partonic channels and integrated over the phase space of the mediator, and any such average produces a mixed state. The second is genuinely quantum: tracing out the invisible particle discards whatever spin entanglement it shares with the $t\bar t$ pair, so that information is lost from the reduced state and QI observables can be sensitive to it. The two $X$ mediator hypotheses probe these mechanisms differently. 
A spin-0 mediator has no spin degrees of freedom to be traced out. At a fixed phase-space point and for fixed initial-state helicities, the reduced $t\bar t$ spin state therefore remains almost pure (with some mixedness appearing for tracing out the color degrees of freedom), and the dominant mixedness comes from the incoherent averages described above. A $Y_0$ can still modify the spin correlations, and hence the entanglement of $\rho_{t\bar t}$. It does so through the modified kinematics of the recoiling pair and through the different Lorentz structure of the $t\bar t Y_0$ production amplitude. A spin-1 mediator acts through both of these effects, and in addition its polarization states can be entangled with the $t\bar t$ spins. Tracing them out then makes the reduced state mixed even at fixed kinematics, which can genuinely degrade its entanglement. 

In this analysis, we focus on the dileptonic $t\bar t$ final state,
characterized by two oppositely charged leptons, two $b$ jets, and
missing transverse momentum. We first assess the expected exclusion
sensitivity to invisible spin-0 and spin-1 mediators by comparing four
complementary observables: the inclusive event yield, the visible
invariant-mass distribution $dN/d\mttvis$, the event-level angular
distribution $dN/dD_{\rm evt}$, and the fiducial spin-correlation
observable $\mathcal D(\mttvis)$. These are treated as alternative
analyses of the same selected event sample, allowing us to quantify the
additional sensitivity carried by kinematic and spin-dependent
information. Expected limits are obtained with profile-likelihood
methods, including systematic and finite-Monte-Carlo uncertainties.

For the exclusion analysis, we find that observables retaining spin
information provide a clear improvement over the inclusive rate and over
purely kinematic information. In particular, the event-level angular
distribution $dN/dD_{\rm evt}$ is more sensitive than
$dN/d\mttvis$, while the fiducial spin-correlation observable
$\mathcal D(\mttvis)$ provides the strongest expected exclusion reach
across the mediator benchmarks considered. The same qualitative pattern
is observed for both spin-0 and spin-1 mediators and remains stable over
a broad range of systematic assumptions. We further show that the spin
information also improves the discrimination between different coupling
structures, highlighting its potential not only for constraining new
physics, but also for characterizing the nature of the underlying
interaction.

This paper is organized as follows. In Sec.~\ref{sec:theo-framework}, we introduce the simplified mediator models and review the spin-density-matrix
formalism and the quantum observables used in the analysis.
In Sec.~\ref{sec:simulation}, we describe the Monte Carlo simulation and
event generation. In Sec.~\ref{sec:analysis-procedure}, we present the event
selection, the extraction of the spin observables, and the statistical
framework used to derive the expected sensitivities. The results for the
exclusion of spin-0 and spin-1 mediators, together with the study of the
discrimination between different coupling structures, are presented in
Sec.~\ref{sec:res}. We conclude in Sec.~\ref{sec:conclusions}.

\section{Theoretical Framework} \label{sec:theo-framework}

In this section we review the formalism used to describe the $\ttbar$ system as a two-qubit state and the associated quantum information observables. We then introduce the simplified DM model used for this analysis and discuss how the presence of an invisible mediator $X$ in $pp \to t\bar{t} + X$ modifies the $\ttbar$ spin density matrix relative to the Standard Model expectation. 

\subsection{Quantum Tomography}\label{subsec:quantum-tom}

The $\ttbar$ system can be treated as a pair of spin-$1/2$ states, i.e.\ a two-qubit system. Such a system is fully described by a $4\times4$ density matrix $\rho$. In the basis of the Pauli matrices $\sigma_i$ ($i=1,2,3$), the spin density matrix can be decomposed as
\begin{equation}
    \rho = \frac{1}{4}\left(\mathbb{1}\otimes\mathbb{1} + \sum_{i} B_i^1\, \sigma_i\otimes\mathbb{1} + \sum_{i} B_i^2\, \mathbb{1}\otimes\sigma_i + \sum_{i,j} C_{ij}\, \sigma_i\otimes\sigma_j \right),
\label{eq:density-matrix}
\end{equation}
where $B_i^1$ and $B_i^2$ are the polarization vectors of qubit 1 and qubit 2, respectively, and $C_{ij}$ is the spin-correlation matrix between them. These quantities have recently been measured by the CMS Collaboration, which simultaneously extracted the full set of 15 polarization and spin-correlation coefficients in $\ttbar$ events~\cite{CMS:2024zkc}.

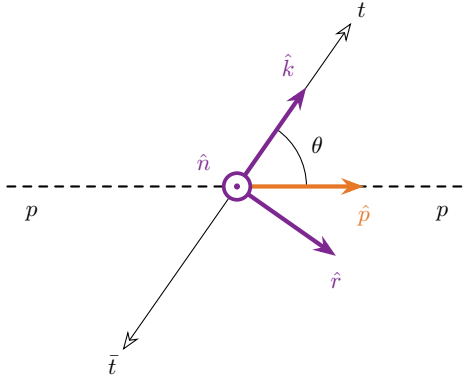
\begin{figure}[t]
  \centering
  \begin{tikzpicture}[scale=0.8, transform shape, line cap=round, >={Stealth[length=3mm,width=2.2mm]}]
    \def\th{55}   % angle theta between k-hat and p-hat (degrees)

    % beam axis
    \draw[dashed, thick] (-3.8,0) -- (3.8,0);
    \node at (-3.4,-0.45) {$p$};
    \node at ( 3.4,-0.45) {$p$};
 
    % t / tbar axis
    \draw[<->, >={Stealth[length=2.5mm,width=1.8mm,open]}]
      ({\th+180}:3.3) -- (\th:3.3);
    \node at ({\th}:3.3)   [above right=-1pt] {$t$};
    \node at ({\th+180}:3.3) [below left=-1pt] {$\bar t$};
 
    % angle theta
    \draw (1.15,0) arc[start angle=0, end angle=\th, radius=1.15];
    \node at ({\th/2}:1.5) {$\theta$};
 
    % p-hat
    \draw[->, vorange, line width=1.6pt] (0,0) -- (2.1,0)
      node[below=4pt] {$\hat{p}$};
 
    % k-hat and r-hat
    \draw[->, vpurple, line width=1.6pt] (0,0) -- (\th:2)
      node[above left=1pt] {$\hat{k}$};
    \draw[->, vpurple, line width=1.6pt] (0,0) -- ({\th-90}:2)
      node[below=3pt] {$\hat{r}$};
 
    % n-hat (out of the page)
    \draw[vpurple, line width=1.4pt, fill=white] (0,0) circle (0.22);
    \fill[vpurple] (0,0) circle (0.05);
    \node[vpurple] at (-0.55,0.4) {$\hat{n}$};
  \end{tikzpicture}
  \caption{Definition of the helicity axes $\{\hat{k},\hat{r},\hat{n}\}$ in the $\ttbar$ rest frame, following the convention of Ref.~\cite{Maltoni2024Quantum}. Figure adapted from Ref.~\cite{Maltoni2024Quantum}.}
    \label{fig:krn_syste}
\end{figure}

The spin density matrix of the $\ttbar$ system, $\rho_{t\bar{t}}$, is not directly observable, since the top and antitop spins cannot be measured directly. However, since the top quark decays before hadronization ($\tau_t \ll \tau_{\rm QCD}$), its spin information is transferred to the angular distributions of its decay products. In the dileptonic channel, the charged leptons act as nearly perfect spin analyzers, allowing $\rho_{t\bar{t}}$ to be reconstructed through a quantum tomography procedure based on the angular distributions of the $\ttbar$ decay products. In this work, we follow the convention adopted in~\cite{Maltoni2024Quantum}, where the axes are defined in the $\ttbar$ rest frame as
\begin{equation}
    \hat{k}=\text{top direction},\quad
    \hat{r}=\frac{\hat{p}-\hat{k}\cos\theta}{\sin\theta},\quad
    \hat{n}=\frac{\hat{p}\times\hat{k}}{\sin\theta}\,,
    \label{eq:hel-basis}
\end{equation}
where $\hat{p}$ is the beam-axis unit vector and $\theta\in[0,\pi/2]$ is the top scattering angle in the $\ttbar$ rest frame, as illustrated in Figure~\ref{fig:krn_syste}.

To extract the angular correlations of the $\ttbar$ decay products, the relevant angles must be evaluated in the rest frame of each top quark, which requires reconstructing the full kinematics of the $\ttbar$ system. In the dileptonic channel, the presence of two neutrinos makes this reconstruction non-trivial but experimentally feasible, as demonstrated by the techniques employed by the ATLAS and CMS Collaborations in spin-correlation measurements~\cite{ATLAS:2023fsd,CMS:2019nrx}. The presence of additional invisible particles, as in the signal and some of the backgrounds considered in this work, introduces further ambiguities in the reconstruction. This issue and possible reconstruction strategies are discussed in Appendix~\ref{app:reco}.

The differential cross section for $\ttbar$ production and subsequent decays, with $a$ and $b$ denoting decay products of the top and antitop quarks, respectively, is given at leading order (LO) by~\cite{Bernreuther2015yna}
\begin{equation}
    \frac{1}{\sigma}\frac{d\sigma}{d(\cos\theta_{ai}\cos\theta_{bj})}
    =-\frac{1+C_{ij}\alpha_a\alpha_b\cos\theta_{ai}\cos\theta_{bj}}{2}
    \log|\cos\theta_{ai}\cos\theta_{bj}|\,,
    \label{eq:diff_theta}
\end{equation}
where $\theta_{ai}$ ($\theta_{bj}$) is the angle between the momentum of $a$ ($b$) and the $i$-th ($j$-th) axis in the top (antitop) rest frame. The parameters $\alpha_a$ and $\alpha_b$ are the spin-analyzing powers of particles $a$ and $b$, respectively, and quantify the correlation between the direction of the emitted particle and the spin of its parent particle. In the SM, light charged leptons act as nearly perfect spin analyzers, with $\alpha_{\ell^{\pm}} \simeq \pm 1$~\cite{ParticleDataGroup:2024cfk}, making the dileptonic channel particularly well suited for studies of top-quark spin correlations.

From Eq.~\eqref{eq:diff_theta}, it follows that the spin-correlation coefficients are given by
\begin{equation}
    C_{ij}=\frac{9}{\alpha_a\alpha_b}
    \left\langle\cos\theta_{ai}\cos\theta_{bj}\right\rangle\,.
    \label{eq:cij}
\end{equation}
Thus, the elements of the spin-correlation matrix can be directly determined from the angular distributions of the decay products.

\subsection{Quantum Observables}\label{subsec:quant_obs}

The spin density matrix $\rho_{t\bar{t}}$, and in particular the spin-correlation matrix $C$, encode the spin-correlation properties of the $\ttbar$ system.  However, its individual entries depend on the choice of the reference axes ($\{\hat{k},\hat{r},\hat{n}\}$ in this paper), making it useful to construct basis-independent quantities that allow for a direct comparison across different conventions. In particular, one can take the trace of $C$ and define the scalar observable
\begin{equation}
    D\equiv \frac{1}{3}\left(C_{kk}+C_{rr}+C_{nn}\right)=\frac{1}{3}\text{Tr}(C). 
    \label{eq:D1}
\end{equation}
An important property of this observable is that $D<-1/3$ is a sufficient condition for the \(\ttbar\) pair to be entangled through the Peres-Horodecki separability criterion \cite{Peres:1996dw,Horodecki:1996nc}. 

The condition $D<-1/3$ is sufficient, but not necessary, for entanglement. $D$ is directly related to the overlap of $\rho_{t\bar t}$ with the spin-singlet state, $\langle\Psi^-|\rho|\Psi^-\rangle=(1-3D)/4$, and no separable two-qubit state can have an overlap larger than $1/2$ with a maximally entangled state. It is not necessary because $D$ only probes the singlet component of $\rho_{t\bar t}$. An entangled state dominated by a different maximally entangled state, such as the triplet Bell states, has $D=+1/3$ and is not detected. For this reason $D$ is a witness of entanglement rather than a complete measure of it. Quantities such as the concurrence make use of the full spin-correlation matrix $C$ and can quantify the entanglement of any two-qubit state. In this work we restrict ourselves to $D$, which is
sufficient for our purposes since we use it as a discriminant between the SM
and SM$+$BSM hypotheses, and not as an entanglement measure.

\subsection{Signal Model}
\label{sec:signal-model}

As signal model, we consider simplified dark matter scenarios in which the SM is extended by a massive spin-0 ($Y_0$) or spin-1 ($Y_1$) mediator~\cite{Abdallah:2015ter}. These states couple to the top quark and, in the underlying dark matter framework, may also couple to a dark sector and decay into a pair of dark matter particles. In the present analysis, however, $Y_{0,1}$ is treated as a stable invisible particle and plays the role of the state $X$ introduced above. The signal process is therefore $pp\to t\bar{t}+Y_{0,1}$, leading to a $\ttbar$ system produced in association with missing transverse momentum.

\begin{figure}[t!]
    \centering
    \includegraphics[width=\linewidth]{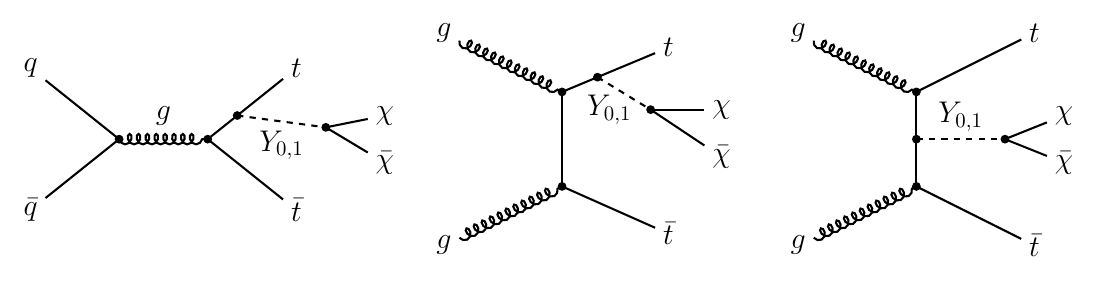}
    \caption{Representative diagrams for the associated production of a spin-0 or spin-1 mediator with a $\ttbar$ pair. The decay $Y_{0,1}\to\chi\bar{\chi}$ is shown for illustration only; in this work the mediator is treated as a stable invisible particle.}
    \label{fig:feynman_diags}
\end{figure}

The relevant interactions are described by
\begin{align}
    \mathcal{L} &\supset \frac{1}{\sqrt{2}}y_t\bar{t}
    \left(g_s+i\,g_p\,\gamma_5\right)t\,Y_0 \,, \nonumber\\
    \mathcal{L} &\supset \bar{t}\gamma^\mu
    \left(g_V+g_A\,\gamma_5\right)t\,Y_{1\mu}\,,
    \label{eq:lagrangian-signal}
\end{align}
where $y_t$ is the top-quark Yukawa coupling, while $g_s$ and $g_p$ ($g_V$ and $g_A$) denote the scalar and pseudoscalar (vector and axial-vector) couplings of $Y_0$ ($Y_1$) to the top quark. Representative production diagrams for $pp\to t\bar{t}+Y_{0,1}$ are shown in Fig.~\ref{fig:feynman_diags}.

Since $Y_{0,1}$ is invisible, it cannot be reconstructed directly. Information on the production process is instead carried by the visible $\ttbar$ system and by the associated missing transverse momentum. In particular, the presence of the invisible mediator can modify the $\ttbar$ spin density matrix in two conceptually distinct ways.
First, the $\ttbar$ system recoils against $Y_{0,1}$, modifying its kinematic distributions, for instance in $m_{t\bar{t}}$ and $p_T(t\bar{t})$, relative to SM $\ttbar$ production. Since $\rho_{t\bar{t}}$ depends on the kinematics of the $\ttbar$ pair, this kinematic reshaping can induce changes in the quantum observables even if the spin density matrix were identical to the SM one at fixed kinematics.
Second, the $t\bar{t}Y_{0,1}$ interaction in Eq.~\eqref{eq:lagrangian-signal} modifies the spin structure of the production amplitude itself. Consequently, $\rho_{t\bar{t}}$ can differ from its SM counterpart even at fixed $\ttbar$ kinematics, reflecting the different Lorentz structure of the scalar, pseudoscalar, vector, and axial-vector interactions.

The resulting deviations in the quantum observable defined in Sec.~\ref{subsec:quant_obs} can therefore arise from both kinematic and dynamical effects. The goal of this work is to quantify these deviations and assess their sensitivity to the signal models considered here, complementing more conventional searches based on missing transverse momentum.
\section{Simulation setup}
\label{sec:simulation}

In this section, we describe the Monte Carlo setup used to simulate the signal and background processes. All samples are generated for proton--proton collisions at a centre-of-mass energy of $\sqrt{s}=14~\mathrm{TeV}$.

The dominant $t\bar t$ background is generated at next-to-leading order (NLO) in QCD using the \texttt{POWHEG BOX} implementation~\cite{Alioli:2010xd} of top-pair production and decay~\cite{Campbell:2014kua}. The subleading $tW$, $t\bar tW$, and $t\bar tZ$ backgrounds are generated at leading order (LO) using \textsc{MadGraph5\_aMC@NLO}~\cite{Alwall:2014hca}. The signal processes,
\begin{equation}
pp\to t\bar tY_0
\qquad\text{and}\qquad
pp\to t\bar tY_1,
\end{equation}
are also generated at LO using \textsc{MadGraph5\_aMC@NLO}, employing the \texttt{DMsimp\_s\_spin0} and \texttt{DMsimp\_s\_spin1} UFO~\cite{Backovic:2015soa} implementations of the simplified models described in Sec.~\ref{sec:signal-model}. The mediators $Y_0$ and $Y_1$ are treated as stable invisible particles and therefore contribute directly to the missing transverse momentum of the event.

We consider mediator masses
\begin{equation}
m_{Y_{0,1}} = 10,\ 25,\ 50,\ 100~\mathrm{GeV}.
\end{equation}
The coupling benchmarks used for the different mediator hypotheses are specified together with the corresponding results in Sec.~\ref{sec:res}.

The top quarks are required to decay in the dileptonic channel,
\begin{equation}
t\bar t\to
b\,\ell^+\nu_\ell\,
\bar b\,\ell'^-\bar\nu_{\ell'},
\qquad \ell,\ell'=e,\mu.
\end{equation}
For the $t\bar t$ sample generated with \texttt{POWHEG BOX}, both production and decay are described at NLO while retaining the spin correlations between the two stages~\cite{Campbell:2014kua}. For the samples generated with \textsc{MadGraph5\_aMC@NLO}, the top-quark and electroweak-boson decays are performed with \textsc{MadSpin}~\cite{Artoisenet:2012st}, preserving the relevant spin correlations. All samples are subsequently interfaced with \textsc{Pythia~8}~\cite{Sjostrand:2014zea} for parton showering and hadronisation.

The targeted final state contains two oppositely charged leptons, two $b$-jets, and missing transverse momentum. In the signal processes, the latter receives contributions both from the neutrinos produced in the top-quark decays and from the invisible mediator. Several SM processes can populate this final state, which we group into three broad categories:
\begin{enumerate}
\item \textbf{Top-quark backgrounds.}
The dominant contribution is dileptonic $t\bar t$ production, which contains two charged leptons, two $b$-quarks, and genuine missing transverse momentum from the two neutrinos. Single-top production in association with a $W$ boson, $tW$, can also enter the signal region when an additional $b$-jet is produced through QCD radiation, for example through gluon splitting into a $b\bar b$ pair.

\item \textbf{Reducible backgrounds.}
This category includes diboson production, $WW$, $WZ$, and $ZZ$, as well as $Z+$jets. These processes may populate the signal region through additional heavy-flavour production, light-jet misidentification, unreconstructed leptons, or instrumental missing transverse momentum.

\item \textbf{Associated top-pair backgrounds.}
This category contains $t\bar tV$ production, with $V=W,Z$. In particular, $t\bar tZ$ production with $Z\to\nu\bar\nu$ constitutes an irreducible background, since it produces the same visible objects as the signal together with additional genuine missing transverse momentum. The $t\bar tW$ process can populate the signal region when two of the three $W$ bosons decay leptonically and the remaining one decays hadronically, or when all three decay leptonically but one charged lepton is not reconstructed or falls outside the detector acceptance.
\end{enumerate}

In this work, we explicitly simulate the $t\bar t$, $tW$, $t\bar tW$, and $t\bar tZ$ backgrounds. The remaining reducible backgrounds are expected to provide only subleading contributions after the event selection and are therefore neglected. Moreover, their residual contribution depends strongly on detector-level effects such as object-identification efficiencies, jet misidentification, and instrumental missing transverse momentum. A reliable estimate of these contributions would therefore require a dedicated detector simulation, which lies beyond the scope of the present phenomenological study. The event selection described in the following section is expected to suppress such backgrounds. The resulting background composition should therefore be interpreted as a phenomenological projection rather than as a complete experimental estimate.

Event yields are evaluated for an integrated luminosity of
\begin{equation}
\mathcal{L}=3000~\mathrm{fb}^{-1}=3~\mathrm{ab}^{-1},
\end{equation}
corresponding to the target dataset of the High-Luminosity LHC at $\sqrt{s}=14~\mathrm{TeV}$~\cite{Cepeda:2019klc}. This choice allows us to assess the sensitivity of the spin-correlation observables in the large-statistics regime expected after the full HL-LHC programme.

\section{Analysis procedure}
\label{sec:analysis-procedure}

As discussed in Sec.~\ref{sec:simulation}, the analysis targets a final
state containing two oppositely charged leptons, at least two
$b$-tagged jets, and missing transverse momentum. In the signal
processes, the missing transverse momentum receives contributions from
both the neutrinos produced in the top-quark decays and the invisible
mediator $Y_{0,1}$.

\subsection{Object reconstruction and event selection}
\label{subsec:event-selection}

Stable final-state particles obtained after parton showering and hadronisation with \textsc{Pythia~8} are used to construct the event objects. Jets are clustered with the anti-\(k_T\) algorithm, as implemented in \textsc{FastJet}~\cite{Cacciari:2011ma}, using a radius parameter \(R=0.4\) and a minimum transverse momentum of \(20~\mathrm{GeV}\). Muons, neutrinos, and the invisible mediator are excluded from the jet clustering. To avoid double counting and ensure lepton isolation, jets within \(\Delta R<0.2\) of an electron are removed, and selected leptons are required to satisfy \(\Delta R(\ell,j)>0.4\) with respect to the nearest surviving jet.

The two highest-$p_T$ isolated lepton candidates are retained, and
events are required to contain an opposite-sign dilepton pair. The
leading and subleading leptons must satisfy 
\begin{equation}
    p_T^{\ell_1}>25~\mathrm{GeV},
    \qquad
    p_T^{\ell_2}>20~\mathrm{GeV}.
    \label{eq:lepton-selection}
\end{equation}
These requirements define a fiducial dilepton selection representative
of the transverse-momentum thresholds commonly employed in
experimental analyses. Since a full detector simulation is not
performed, they should be interpreted as an approximation to an
experimentally accessible lepton selection rather than as a detailed
description of trigger and reconstruction efficiencies.

The flavour of each jet is assigned by matching it to a hard-process bottom or charm quark within \(\Delta R<0.4\). A simplified parametric flavour-tagging model is then applied, with a \(b\)-tagging efficiency of \(\epsilon_b=0.70\) and mistag probabilities of \(\epsilon_c=0.15\) and \(\epsilon_{\mathrm{light}}=0.01\) for charm and light-flavour jets, respectively.

Events are required to contain at least two $b$-tagged jets satisfying
\begin{equation}
    p_T^b>30~\mathrm{GeV},
    \qquad
    |\eta_b|<2.5.
    \label{eq:bjet-selection}
\end{equation}
If more than two tagged jets are present, the two with the highest
transverse momentum are retained for the construction of the analysis
observables. These requirements select sufficiently energetic jets
within the central detector region, where $b$-jet identification is
experimentally feasible.

The missing transverse momentum is defined at generator level as the
vector sum of the transverse momenta of all invisible final-state
particles,
\begin{equation}
    \vec p_T^{\,\mathrm{miss}}
    =
    \sum_{\nu}\vec p_{T,\nu}
    +\vec p_{T,Y},
    \label{eq:met-definition}
\end{equation}
where the sum runs over all final-state neutrinos and the last term is
present only for signal samples containing an invisible mediator.

We also define the visible invariant mass of the $t\bar t$ system as
\begin{equation}
    M_{t\bar t}^{\rm vis}
    \equiv
    \sqrt{
    \left(
        p_{\ell_1}+p_{\ell_2}+p_{b_1}+p_{b_2}
    \right)^2
    },
    \label{eq:mttvis-definition}
\end{equation}
where $p_{\ell_{1,2}}$ and $p_{b_{1,2}}$ denote the four-momenta of the two selected leptons and $b$-tagged jets, respectively. It therefore corresponds to the invariant mass reconstructed from the visible decay products of the dileptonic $t\bar t$ system, without attempting to reconstruct the neutrino momenta.

Among the kinematic variables considered in the analysis, the
stransverse mass $m_{T2}^{\ell\ell}$~\cite{Lester:1999tx,Barr:2003rg}
is constructed from the transverse momenta of the two selected leptons
and from $\vec p_T^{\,\mathrm{miss}}$. Assuming massless invisible test
particles, it is defined as
\begin{equation}
    m_{T2}^{\ell\ell}
    =
    \min_{\vec q_{T,1}+\vec q_{T,2}=\vec p_T^{\,\mathrm{miss}}}
    \left[
    \max\left(
    m_T(\vec p_T^{\,\ell_1},\vec q_{T,1}),
    m_T(\vec p_T^{\,\ell_2},\vec q_{T,2})
    \right)
    \right],
    \label{eq:mt2-definition}
\end{equation}
where
\begin{equation}
    m_T^2(\vec p_T,\vec q_T)
    =
    2\,p_T q_T
    \left[1-\cos\Delta\phi(\vec p_T,\vec q_T)\right].
\end{equation}
Here, $\Delta\phi(\vec p_T,\vec q_T)$ denotes the azimuthal separation between the two transverse-momentum vectors.
For dileptonic $t\bar t$ events, $m_{T2}^{\ell\ell}$ is bounded by the
$W$-boson mass at parton level, up to finite-width, radiation, and
reconstruction effects, while additional invisible particles can
populate the region above this endpoint.

\begin{figure}[t]
    \centering
    \includegraphics[width=\linewidth]
        {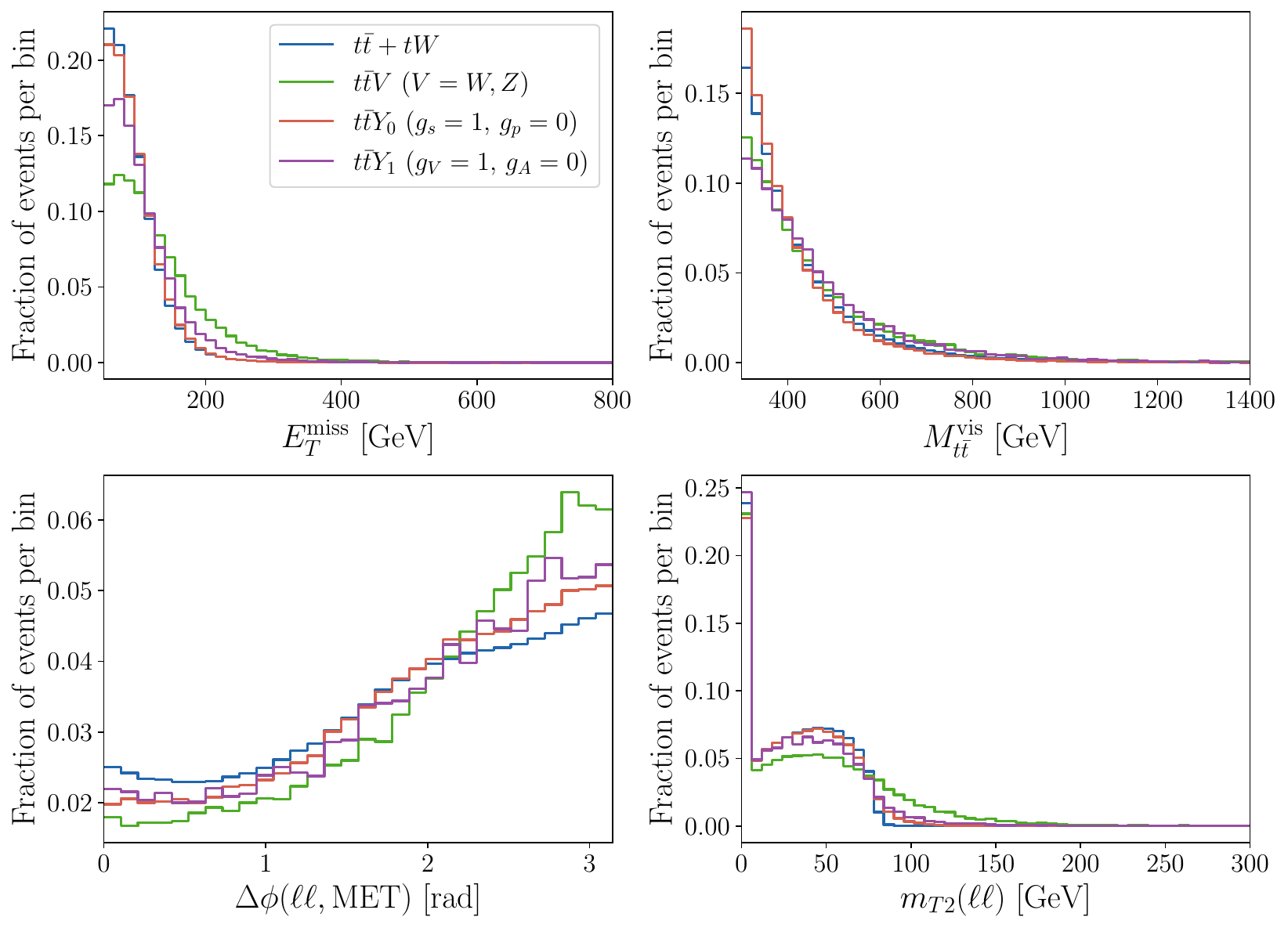}
    \caption{
        Normalized distributions of $E_T^{\rm miss}$,
        $M_{t\bar t}^{\rm vis}$,
        $\Delta\phi(\ell\ell,\vec p_T^{\,\rm miss})$, and
        $m_{T2}^{\ell\ell}$ after the baseline object selection and
        before the signal-enhancing kinematic cuts. The top-quark
        background contains the $t\bar t$ and $tW$ processes, while
        the $t\bar tV$ contribution contains $t\bar tW$ and
        $t\bar tZ$. The signal benchmarks correspond to
        $t\bar tY_0$ and $t\bar tY_1$ production with
        $m_{Y_{0,1}}=10~\mathrm{GeV}$.
    }
    \label{fig:pre-cut-distributions}
\end{figure}

Figure~\ref{fig:pre-cut-distributions} shows the normalized
distributions of selected kinematic observables after the baseline
object selection and before the signal-enhancing kinematic
requirements. The SM processes are grouped into a top-quark background,
comprising $t\bar t$ and $tW$, and an associated top-pair background,
comprising $t\bar tW$ and $t\bar tZ$. The signal benchmarks shown
correspond to $t\bar tY_0$ and $t\bar tY_1$, with mediator masses
$m_{Y_0}=m_{Y_1}=10~\mathrm{GeV}$. The azimuthal variable $\Delta\phi(\ell\ell,\vec p_T^{\,\mathrm{miss}})$ is defined as the separation in the transverse plane between the dilepton momentum,
$\vec p_T^{\,\ell\ell}=\vec p_T^{\,\ell_1}+\vec p_T^{\,\ell_2}$,
and the missing transverse momentum.

The $E_T^{\rm miss}$ and $m_{T2}^{\ell\ell}$ distributions provide useful discrimination between the dominant top-quark background and the signal. The signal tends to exhibit a harder missing-transverse-momentum spectrum due to the recoil against the invisible mediator, although a substantial overlap with the SM background remains. A more pronounced separation is provided by $m_{T2}^{\ell\ell}$. In dileptonic $t\bar t$ events, where the only invisible particles are the two neutrinos from the $W$-boson decay chains, the distribution is concentrated below the $W$-boson mass. Signal events and backgrounds containing additional invisible particles can instead populate the region above this endpoint. The distributions of $M_{t\bar t}^{\rm vis}$ and $\Delta\phi(\ell\ell,\vec p_T^{\,\rm miss})$ provide comparatively less discrimination and are therefore not used to define the event selection.

Motivated by these distributions, we impose the simultaneous requirements
\begin{equation}
    \Etmiss>100~\mathrm{GeV},
    \qquad
    \mttwo^{\ell\ell}>90~\mathrm{GeV}.
    \label{eq:analysis-cuts}
\end{equation}

The $m_{T2}^{\ell\ell}$ requirement is particularly effective in suppressing the dominant dileptonic $t\bar t$ contribution. In the idealized case of on-shell $W$ bosons and perfect reconstruction, the $m_{T2}^{\ell\ell}$ distribution has a kinematic endpoint at the $W$-boson mass. Parton showering, finite-width effects, and the event selection smear this endpoint, but the bulk of the $t\bar t$ contribution remains concentrated at lower values of $m_{T2}^{\ell\ell}$. Signal events containing an invisible mediator, as well as SM backgrounds with additional invisible particles such as $t\bar tZ$ with $Z\to\nu\bar\nu$, can instead populate the region above the endpoint. The requirement on $E_T^{\rm miss}$ provides an additional selection on events characterized by sizeable invisible recoil.

The cumulative impact of these requirements is reported in Table~\ref{tab:cutflow}. For the signal benchmarks, we fix the scalar coupling to $g_s=1$. Since the spin-0 and spin-1 interactions in Eq.~\eqref{eq:lagrangian-signal} use different coupling normalizations, we define the reference vector coupling by matching the effective top-mediator interaction strength,
\begin{equation}
    g_V^{\rm ref}=\frac{y_t}{\sqrt{2}}\,g_s.
    \label{eq:vector-reference-coupling}
\end{equation}
For $g_s=1$, this gives $g_V^{\rm ref}\simeq0.70$.

The $E_T^{\rm miss}$ requirement reduces the event yields of both the signal and the dominant top-quark background by comparable factors, and therefore provides only limited discrimination on its own. The subsequent $m_{T2}^{\ell\ell}$ requirement is considerably more effective in suppressing the dileptonic $t\bar t$ background, while retaining a larger fraction of the signal and of backgrounds containing additional invisible particles.

\begin{table}[t]
\centering
\footnotesize
\begin{tabular}{lrrr}
\toprule
Process
& Before kinematic cuts
& After $\Etmiss>100~\mathrm{GeV}$
& After $\mttwo^{\ell\ell}>90~\mathrm{GeV}$ \\
\midrule
$t\bar t+tW$
& $32{\ }834{\ }611$
& $7{\ }770{\ }591$ $(23.7\%)$
& $13{\ }707$ $(0.04\%)$ \\

$t\bar tV$
& $4{\ }547$
& $2{\ }268$ $(48.0\%)$
& $593$ $(12.5\%)$ \\

$t\bar tY_0$
& $515{\ }051$
& $130{\ }833$ $(25.4\%)$
& $6{\ }766$ $(1.3\%)$ \\

$t\bar tY_1$
& $510{\ }976$
& $172{\ }884$ $(33.8\%)$
& $20{\ }578$ $(4.0\%)$ \\
\bottomrule
\end{tabular}
\caption{
Cumulative cut-flow for the main SM backgrounds and the
$t\bar tY_0$ and $t\bar tY_1$ signal benchmarks after the
baseline object selection. The entries correspond to expected
event yields normalized to $\mathcal{L}=3~\mathrm{ab}^{-1}$. The $t\bar tV$ category
contains the $t\bar tW$ and $t\bar tZ$ processes. The signal benchmarks correspond to $m_{Y_0}=m_{Y_1}=10~\mathrm{GeV}$, with $g_s=1$ and $g_V=0.70$.}
\label{tab:cutflow}
\end{table}

A tighter requirement on $E_T^{\rm miss}$ or
$m_{T2}^{\ell\ell}$ could further suppress the dominant top-quark
background at generator level. We nevertheless adopt the comparatively
moderate thresholds in Eq.~\eqref{eq:analysis-cuts}. Since the present
analysis does not include a full detector simulation, effects such as
the experimental resolution of the visible objects,
detector-induced missing momentum, and the resulting migration of
$t\bar t$ events across the nominal $m_{T2}^{\ell\ell}$ endpoint are
not modeled. Optimizing more aggressive thresholds using the
unsmeared distributions could therefore overestimate the achievable
background rejection. The selected working point is consequently
intended as a conservative choice appropriate for the scope of this
study.

\begin{figure}[t]
\centering

\begin{minipage}[t]{0.48\textwidth}
    \centering
    \includegraphics[width=\linewidth]
        {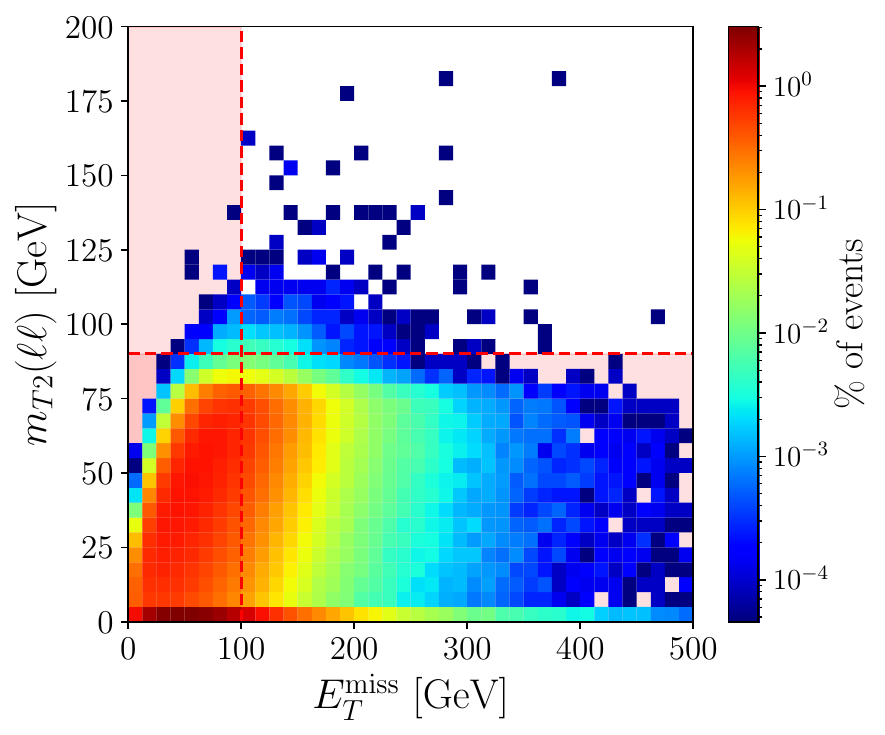}
    \par\vspace{2pt}
    \small (a) $t\bar t+tW$
\end{minipage}
\hfill
\begin{minipage}[t]{0.48\textwidth}
    \centering
    \includegraphics[width=\linewidth]
        {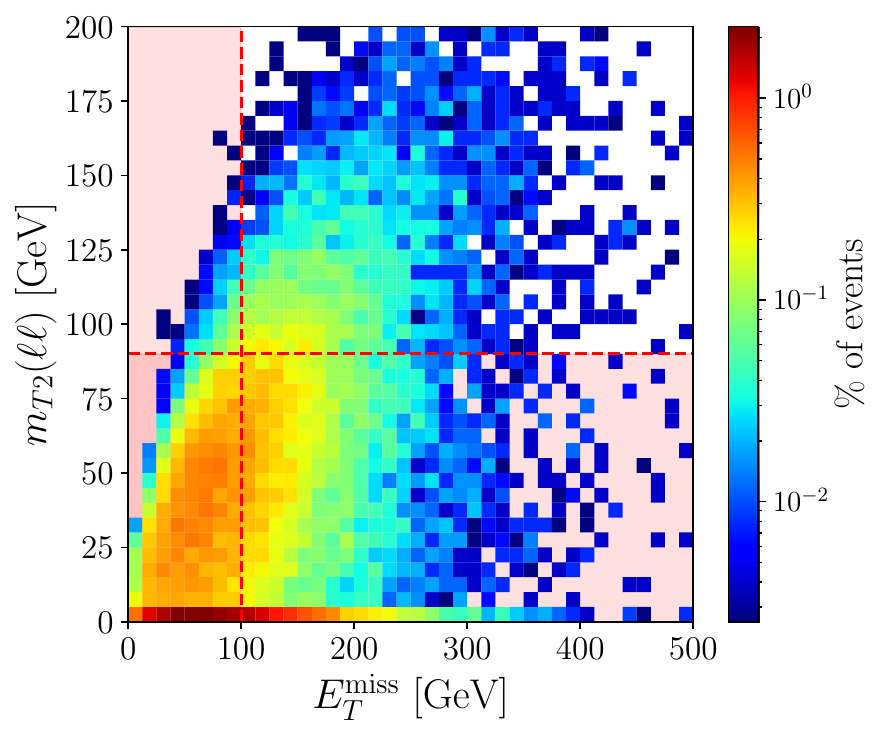}
    \par\vspace{2pt}
    \small (b) $t\bar tV$, $V=W,Z$
\end{minipage}

\vspace{0.8em}

\begin{minipage}[t]{0.48\textwidth}
    \centering
    \includegraphics[width=\linewidth]
        {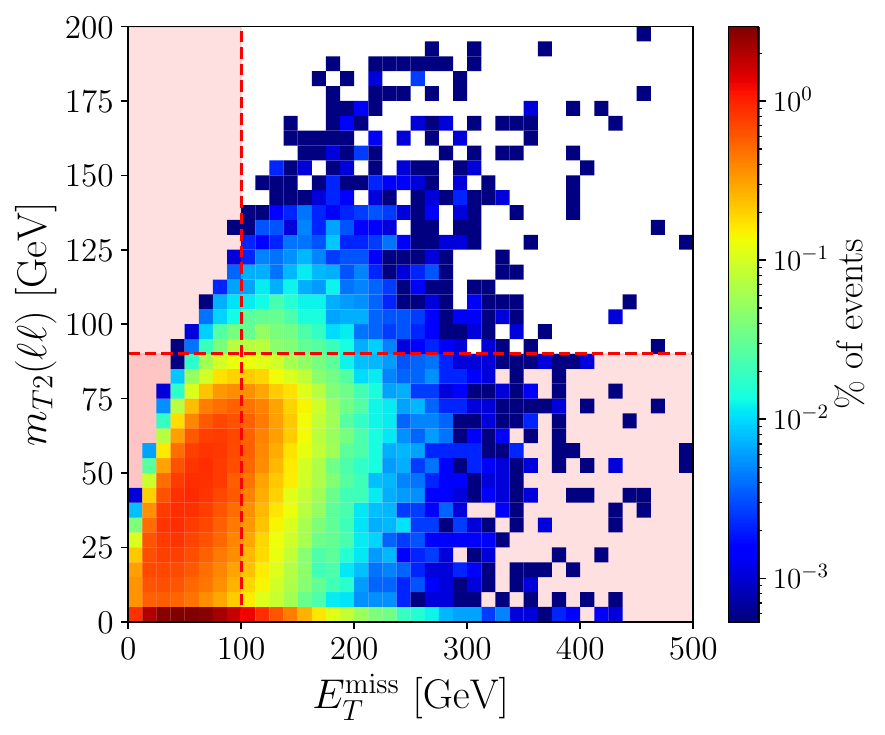}
    \par\vspace{2pt}
    \small (c) $t\bar tY_0$
\end{minipage}
\hfill
\begin{minipage}[t]{0.48\textwidth}
    \centering
    \includegraphics[width=\linewidth]
        {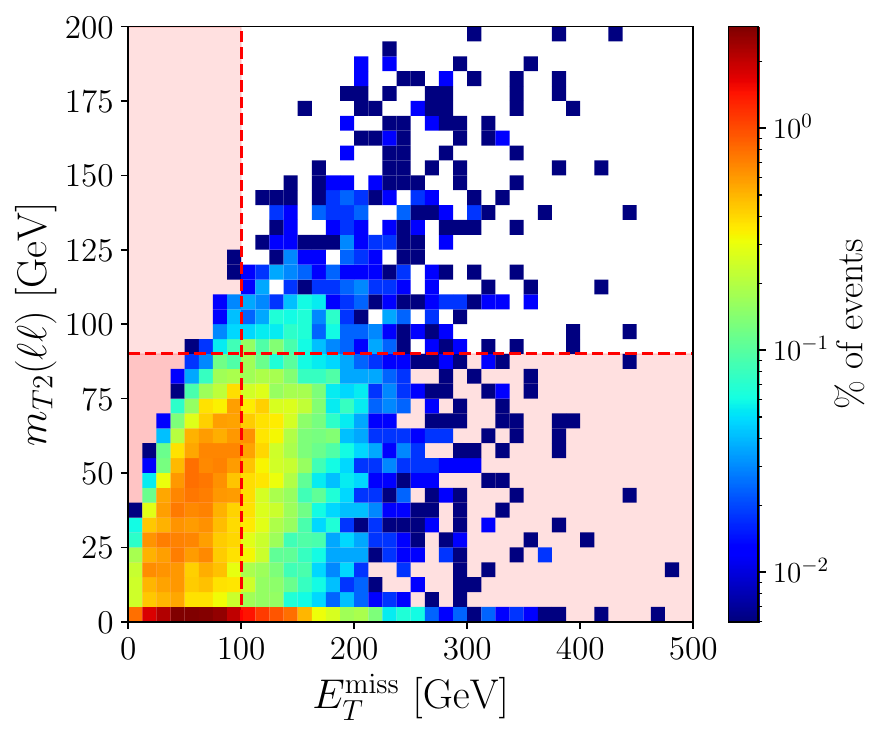}
    \par\vspace{2pt}
    \small (d) $t\bar tY_1$
\end{minipage}

\caption{
    Normalized event distributions in the
    $(E_T^{\rm miss},m_{T2}^{\ell\ell})$ plane after the baseline
    object selection for (a) the top-quark background,
    (b) the associated $t\bar tV$ background,
    (c) the $t\bar tY_0$ signal benchmark, and
    (d) the $t\bar tY_1$ signal benchmark.
    The signal benchmarks correspond to
    $m_{Y_0}=m_{Y_1}=10~\mathrm{GeV}$, with
    $g_s=1$ and $g_V=0.70$.
    The dashed lines indicate the requirements
    $E_T^{\rm miss}>100~\mathrm{GeV}$ and
    $m_{T2}^{\ell\ell}>90~\mathrm{GeV}$; the selected signal
    region lies above and to the right of these thresholds.
}
\label{fig:met-mt2-heatmaps}

\end{figure}

The resulting two-dimensional distributions in the
$(E_T^{\rm miss},m_{T2}^{\ell\ell})$ plane are shown in
Fig.~\ref{fig:met-mt2-heatmaps} for the top-quark background, the
associated top-pair background, and the $t\bar tY_0$ and $t\bar tY_1$
signal benchmarks. The selected signal region, defined by the simultaneous
requirements $E_T^{\rm miss}>100~\mathrm{GeV}$ and
$m_{T2}^{\ell\ell}>90~\mathrm{GeV}$, is primarily designed to suppress the
dominant dileptonic $t\bar t$ background while retaining a sizeable fraction
of the signal. Backgrounds containing additional invisible particles,
most notably $t\bar tZ$ with $Z\to\nu\bar\nu$, populate a similar region of
phase space and are therefore remain as an irreducible component of the selected sample.

\subsection{Extraction of spin observables}
\label{subsec:spin-extraction}

Having defined the event selection, we turn to the extraction of the
spin-correlation observables. The construction of the helicity frame
$\{\hat k,\hat r,\hat n\}$ requires knowledge of the top- and
antitop-quark four-momenta and, consequently, the reconstruction of the
$t\bar t$ rest frame.

For the dileptonic $t\bar t$ process, the two unmeasured neutrino
momenta can in principle be reconstructed by combining the measured
missing transverse momentum with the on-shell constraints associated
with the two $W$ bosons and the two top quark. Although the
resulting nonlinear system can admit multiple discrete solutions,
several reconstruction techniques have been developed and employed in
experimental measurements.

The situation is qualitatively different for processes containing an
additional invisible system,
\begin{equation}
    pp\to t\bar tX,
    \qquad
    X=Y_0,Y_1,Z(\to\nu\bar\nu), W(\to\nu\ell).
\end{equation}
In these events, the missing transverse momentum receives contributions
from both neutrinos produced in the top-quark decay chains and from the
additional invisible system:
\begin{equation}
    \vec p_T^{\,\mathrm{miss}}
    =
    \vec p_{T,\nu}
    +
    \vec p_{T,\bar\nu}
    +
    \vec p_{T,X}.
    \label{eq:met-ttx-reconstruction}
\end{equation}
The standard dileptonic $t\bar t$ reconstruction is therefore
underconstrained and does not provide a unique event-by-event
determination of the top and antitop momenta. This affects both the
signal samples and the irreducible $t\bar tZ$ background with
$Z\to\nu\bar\nu$.

In the present phenomenological analysis, we isolate the intrinsic
sensitivity of the spin observables by taking the top- and
antitop-quark four-momenta directly from the generator-level
hard-process record. These momenta are used to construct the
$t\bar t$ rest frame and the helicity axes, while the charged-lepton
momenta entering the angular observables are taken after parton
showering and object reconstruction. The resulting procedure should
therefore be understood as a truth-assisted reconstruction rather than
as a complete experimental reconstruction strategy. The kinematic
reconstruction problem, together with possible approximate solutions, is discussed in more
detail in Appendix~\ref{app:reco}.

A separate consideration applies to the $tW$ background. Since this
process contains only one top quark, a physical $t\bar t$ rest frame
and the corresponding helicity basis cannot be defined. The $tW$
sample is therefore included in the event selection, in the kinematic
distributions, and in the rate-based analysis, but it is not included
in the extraction of the $t\bar t$ spin-correlation coefficients.
After the full event selection, its expected contribution is
numerically very small compared with the dominant dileptonic
$t\bar t$ background, so that neglecting it in the spin-observable
analysis has a negligible impact on the combined SM prediction.

For each event passing all cuts, the helicity frame
$\{\hat k,\hat r,\hat n\}$ is constructed as described in
Sec.~\ref{subsec:quantum-tom}. In the dileptonic channel considered
here, we identify $a=\ell^+$ and $b=\ell^-$. For each helicity axis
$i\in\{k,r,n\}$, we define $\theta_{\ell^+ i}$ as the angle between
the momentum of the positively charged lepton, evaluated in the top-quark
rest frame, and the $i$-th helicity axis. Analogously,
$\theta_{\ell^- i}$ denotes the angle between the negatively charged
lepton momentum in the antitop-quark rest frame and the corresponding
$i$-th axis.

Events are binned according to their visible invariant mass
$\mttvis$, and the effective spin-correlation coefficients are
extracted in each bin as
\begin{equation}
    \mathcal{S}_{ij}
    \equiv \frac{9}{\alpha_a\alpha_b}
    \langle\cos\theta_{ai}\cos\theta_{bj}\rangle_{\mathcal{F}}\,,
    \label{eq:cij-eff}
\end{equation}
where the average is taken over the events in the bin belonging to the
fiducial region $\mathcal F$ defined by the object selection and the
kinematic requirements in Eq.~\eqref{eq:analysis-cuts}.

In the absence of kinematic cuts, $\mathcal S_{ij}$ reduces to the
corresponding spin-correlation coefficient $C_{ij}$ of the
$t\bar t$ production density matrix. After the fiducial selection,
however, the extracted moments receive acceptance-dependent
distortions and do not necessarily admit the same direct
density-matrix interpretation as the inclusive coefficients. We
therefore define
\begin{equation}
    \Delta\mathcal{S}_{ij}
    =
    \mathcal{S}_{ij}^{\rm SM+BSM}
    -
    \mathcal{S}_{ij}^{\rm SM}
\end{equation}
as a fiducial discriminant between the two hypotheses. Both terms are
evaluated using the same reconstruction, selection, and binning
procedure. The acceptance effects are thus consistently included in
the definition of the observable, although they do not in general
cancel exactly between the SM and BSM hypotheses.

The statistical uncertainty on $\mathcal{S}_{ij}$ in each bin is
estimated as
\begin{equation}
    \delta\mathcal{S}_{ij}
    = \frac{9}{|\alpha_a\alpha_b|}
    \frac{\sigma(\cos\theta_{ai}\cos\theta_{bj})}{\sqrt{N}}\,,
    \label{eq:dcij}
\end{equation}
where $\sigma(\cos\theta_{ai}\cos\theta_{bj})$ is the standard
deviation of the corresponding angular product over the events in the
bin, and $N$ is the expected number of physical events.

The spin-correlation coefficients for a sample containing several
processes are constructed as event-yield-weighted averages:
\begin{equation}
    \mathcal{S}_{ij}^{\rm tot}
    = \frac{\sum_s N_s\,\mathcal{S}_{ij}^s}
           {\sum_s N_s}\,,
    \label{eq:cij-tot}
\end{equation}
where the sum runs over the processes $s$ included in the sample and
$N_s$ denotes the corresponding expected event yield in the given
$\mttvis$ bin. The same prescription is used to construct the SM and
SM-plus-signal hypotheses.

From the spin-correlation matrix $\mathcal{S}$, we also define the
scalar observable
\begin{equation}
    \mathcal{D} \equiv \frac{1}{3}\,\mathrm{Tr}(\mathcal{S})
    = \frac{1}{3}(\mathcal{S}_{kk} + \mathcal{S}_{rr}
    + \mathcal{S}_{nn})\,,
    \label{eq:D_cut}
\end{equation}
which in the absence of kinematic cuts coincides with the standard
entanglement marker $D$ from Eq.~\eqref{eq:D1}.

\subsection{Analysis observables}
\label{subsec:analysis-observables}

To assess the information carried by the spin correlations relative to
more conventional observables, we consider four alternative analysis
strategies. The first uses only the total number of selected events,
$N_{\rm tot}$, and therefore exploits exclusively the overall signal
rate. The second additionally uses the shape of the visible
invariant-mass distribution,
\begin{equation}
    \frac{dN}{d\mttvis}.
\end{equation}
This retains kinematic information on the selected $\ttbar$ system
without explicitly using the angular correlations of its decay
products.

We then consider observables constructed from the angular information
introduced in Sec.~\ref{subsec:spin-extraction}. For each event, we
define
\begin{equation}
    D_{\rm evt}\equiv -3\cos\varphi,
    \qquad
    \cos\varphi
    \equiv
    \sum_{i=k,r,n}
    \cos\theta_{\ell^+ i}\cos\theta_{\ell^- i}.
    \label{eq:D-event}
\end{equation}
The distribution
\begin{equation}
    \frac{dN}{dD_{\rm evt}}
\end{equation}
can be treated as an ordinary event-count distribution and probes
differences in the angular structure of the signal and background
samples.

The event-level quantity $D_{\rm evt}$ should not be interpreted as an
event-by-event entanglement observable. Its ensemble average is instead
directly related to the fiducial spin-correlation observable introduced
in Eq.~\eqref{eq:D_cut}. In a given $\mttvis$ bin $b$,
\begin{equation}
    \mathcal D_b
    =
    \left\langle D_{\rm evt}\right\rangle_{\mathcal F,b}
    =
    -3\left\langle\cos\varphi\right\rangle_{\mathcal F,b},
    \label{eq:D-binned}
\end{equation}
where the average is evaluated over the events satisfying the complete
fiducial selection. Without binning in $\mttvis$, $\mathcal D$ is
therefore simply the first moment of the $D_{\rm evt}$ distribution,
\begin{equation}
    \mathcal D
    =
    \frac{1}{N_{\rm tot}}
    \int dD_{\rm evt}\,
    D_{\rm evt}\,
    \frac{dN}{dD_{\rm evt}}.
    \label{eq:D-first-moment}
\end{equation}
Thus, the inclusive $\mathcal D$ is a compression of the full angular
distribution, retaining only its first moment.

Similarly, $\mathcal D(\mttvis)$ corresponds to the first moment of
the $D_{\rm evt}$ distribution evaluated separately in each
$\mttvis$ bin. It therefore retains information on the dependence of
the spin correlation on the event kinematics that is lost in the
inclusive $dN/dD_{\rm evt}$ distribution.

The four observables used in the sensitivity analysis are therefore
\begin{equation}
    N_{\rm tot},
    \qquad
    \frac{dN}{d\mttvis},
    \qquad
    \frac{dN}{dD_{\rm evt}},
    \qquad
    \mathcal D(\mttvis).
    \label{eq:analysis-observables}
\end{equation}
They are treated as alternative summaries of the same selected event
sample and their likelihoods are not combined.

A natural extension would be to use the full joint distribution
\begin{equation}
    \frac{d^2N}{d\mttvis\,dD_{\rm evt}},
\end{equation}
which contains both one-dimensional distributions as well as the
correlation between $\mttvis$ and $D_{\rm evt}$. In particular,
$\mathcal D(\mttvis)$ can be obtained as the first moment of this
distribution in $D_{\rm evt}$ within each mass bin. A two-dimensional
analysis would therefore retain more differential information.
We do not include this observable in the present sensitivity study,
however. In practice, a two-dimensional template requires a larger
number of bins and correspondingly more shape and finite-Monte-Carlo
nuisance parameters. Its performance therefore becomes more sensitive
to the choice of binning, limited template statistics, and the assumed
correlations of the systematic uncertainties. A robust treatment would
require a more detailed optimization of the two-dimensional likelihood,
which we leave for future work.

Finally, because the fiducial selection can distort the underlying
angular distributions, $\mathcal D$ does not necessarily retain the
direct density-matrix interpretation of the inclusive entanglement
marker $D$ introduced in Sec.~\ref{subsec:quant_obs}. Throughout the
analysis it is therefore interpreted as a fiducial spin-correlation
observable.
\subsection{Statistical methodology}
\label{subsec:stat-methodology}

The sensitivity of the observables introduced in
Sec.~\ref{subsec:analysis-observables} is evaluated using
profile-likelihood analyses. The inclusive yield, $dN/d\mttvis$, and
$dN/dD_{\rm evt}$ are described using Poisson likelihoods, with the
inclusive analysis corresponding to a single bin. For a binned
observable, the expected event yield in bin $b$ is
\begin{equation}
    \lambda_b(\mu,\boldsymbol{\theta})
    =
    B_b(\boldsymbol{\theta})
    +
    \mu S_b(\boldsymbol{\theta}),
    \label{eq:expected-yield}
\end{equation}
where $B_b$ and $S_b$ are the expected background and benchmark signal
yields. The signal strength $\mu$ is the parameter of interest, with
$\mu=0$ corresponding to the SM-only hypothesis and $\mu=1$ to the
reference signal benchmark. The nuisance parameters
$\boldsymbol{\theta}$ account for systematic and Monte Carlo
statistical uncertainties.

The corresponding likelihood is
\begin{equation}
    \mathcal L_N(\mu,\boldsymbol{\theta})
    =
    \prod_b
    \operatorname{Pois}
    \left(
        n_b\mid\lambda_b(\mu,\boldsymbol{\theta})
    \right)
    \prod_k \pi_k(\theta_k),
    \label{eq:likelihood}
\end{equation}
where $\pi_k$ denotes the constraint associated with nuisance parameter
$\theta_k$. Median expected sensitivities are evaluated using the
background-only Asimov dataset, $n_b^{\rm A}=B_b^0$.

The overall background-normalization uncertainty is described by
\begin{equation}
    B_b(\theta_{\rm norm})
    =
    B_b^0
    \left(
        1+f_{\rm norm}\theta_{\rm norm}
    \right),
    \qquad
    \theta_{\rm norm}\sim\mathcal N(0,1),
    \label{eq:background-normalisation}
\end{equation}
with a common nuisance parameter across all bins. The impact of this
uncertainty is assessed by scanning over $f_{\rm norm}$ in
Sec.~\ref{subsec:inclusive-reach}.

For the differential event-yield analyses, an additional shape-only
uncertainty is introduced through independent bin-deformation
parameters $\theta_{{\rm shape},b}$. We use exponential morphing,
normalized such that the total background yield remains unchanged,
\begin{equation}
    \widetilde B_b
    =
    B_b
    \exp\left(
        f_{\rm shape}\theta_{{\rm shape},b}
    \right)
    \frac{
        \sum_c B_c
    }{
        \sum_c
        B_c
        \exp\left(
            f_{\rm shape}\theta_{{\rm shape},c}
        \right)
    },
    \qquad
    \theta_{{\rm shape},b}\sim\mathcal N(0,1).
    \label{eq:shape-morphing}
\end{equation}
The impact of the shape uncertainty is studied by scanning over
$f_{\rm shape}$ in Sec.~\ref{subsec:differential-sensitivity}.

Finite Monte Carlo statistics are included through positive bin-by-bin
scaling parameters,
\begin{equation}
    B_{p,b}\to\gamma_{p,b}B_{p,b},
    \qquad
    S_b\to\gamma_{S,b}S_b,
\end{equation}
where $p$ labels the individual background processes. The scaling
parameters are constrained using auxiliary Poisson terms following the
Barlow--Beeston approach~\cite{Barlow:1993dm}.

The fiducial spin correlations $\mathcal D_b$ are instead compared
using a multivariate Gaussian likelihood. In the presence of a signal,
the predicted value in bin $b$ is the yield-weighted combination
\begin{equation}
    \mathcal D_b^{\rm pred}
    =
    \frac{
        B_b(\boldsymbol{\theta})\,\mathcal D_b^B
        +
        \mu S_b(\boldsymbol{\theta})\,\mathcal D_b^S
    }{
        B_b(\boldsymbol{\theta})
        +
        \mu S_b(\boldsymbol{\theta})
    }.
    \label{eq:D-prediction}
\end{equation}
The background-normalization uncertainty therefore affects the
spin-correlation analysis through the relative signal and background
contributions.

The statistical uncertainty on $\mathcal D_b$ is obtained directly
from the event-level angular distribution,
\begin{equation}
    \delta\mathcal D_b^{\rm stat}
    =
    3\,
    \frac{
        \sigma_b(\cos\varphi)
    }{
        \sqrt{N_b}
    },
    \label{eq:D-stat-uncertainty}
\end{equation}
where $\sigma_b(\cos\varphi)$ is the standard deviation of
$\cos\varphi$ in bin $b$. Finite Monte Carlo uncertainties on both the
angular means and the predicted yields are included separately.

The Gaussian likelihood is
\begin{equation}
    -2\log\mathcal L_{\mathcal D}
    =
    \left(
        \boldsymbol{\mathcal D}^{\rm obs}
        -
        \boldsymbol{\mathcal D}^{\rm pred}
    \right)^T
    V_{\mathcal D}^{-1}
    \left(
        \boldsymbol{\mathcal D}^{\rm obs}
        -
        \boldsymbol{\mathcal D}^{\rm pred}
    \right)
    -
    2\sum_k\log\pi_k(\theta_k),
    \label{eq:D-likelihood}
\end{equation}
where $V_{\mathcal D}$ includes the statistical, finite Monte Carlo,
and systematic uncertainties. Following
Ref.~\cite{Maltoni2024Quantum}, we take an absolute systematic
uncertainty
\begin{equation}
    \delta\mathcal D_{\rm syst}=0.015,
    \label{eq:D-systematic}
\end{equation}
treated as independent between $\mttvis$ bins,
\begin{equation}
    \left(V_{\mathcal D}^{\rm syst}\right)_{bb'}
    =
    \left(\delta\mathcal D_{\rm syst}\right)^2
    \delta_{bb'}.
    \label{eq:D-systematic-covariance}
\end{equation}
Its impact is assessed by considering more conservative values in
Sec.~\ref{subsec:fiducial-spin-sensitivity}.

Since the four observables are constructed from the same selected
events, they are treated as alternative analyses rather than combined
into a single likelihood. Their comparison is used to assess the
additional sensitivity carried by kinematic and spin-dependent
information.

For each analysis, the expected exclusion limit is obtained using the
one-sided profile-likelihood test statistic for an upper limit,
\begin{equation}
 q_\mu
=
\begin{cases}
-2\log
\dfrac{
\mathcal L
\left(
\mu,\hat{\hat{\boldsymbol{\theta}}}_{\mu}
\right)
}{
\mathcal L
\left(
\hat\mu,\hat{\boldsymbol{\theta}}
\right)
},
& 0\leq\hat\mu\leq\mu,
\\[2ex]
-2\log
\dfrac{
\mathcal L
\left(
\mu,\hat{\hat{\boldsymbol{\theta}}}_{\mu}
\right)
}{
\mathcal L
\left(
0,\hat{\hat{\boldsymbol{\theta}}}_{0}
\right)
},
& \hat\mu<0,
\\[2ex]
0,
& \hat\mu>\mu,
\end{cases}
\label{eq:profile-likelihood-ratio}
\end{equation}
where $\hat{\hat{\boldsymbol{\theta}}}_{\mu}$ and
$(\hat\mu,\hat{\boldsymbol{\theta}})$ denote the conditional and
unconditional maximum-likelihood estimators, respectively. The physical
signal strength is restricted to $\mu\geq0$. The median expected limit is
obtained using the asymptotic approximation and the background-only Asimov
dataset~\cite{Cowan:2010js}.

The expected 95\% $CL_s$ upper limit on the signal strength is defined by
\begin{equation}
CL_s(\mu_{\rm lim})=0.05.
\label{eq:cls-threshold}
\end{equation}

Finally, the signal strength can be translated into a limit on the
mediator coupling. Since the signal cross section scales quadratically
with the overall coupling normalization,
\begin{equation}
    \sigma(g)
    =
    \sigma(g_{\rm ref})
    \left(
        \frac{g}{g_{\rm ref}}
    \right)^2,
    \label{eq:coupling-scaling}
\end{equation}
we have
\begin{equation}
    \mu(g)
    =
    \left(
        \frac{g}{g_{\rm ref}}
    \right)^2,
    \qquad
    g_{\rm lim}
    =
    g_{\rm ref}\sqrt{\mu_{\rm lim}}.
    \label{eq:mu-to-coupling}
\end{equation}

\section{Results}
\label{sec:res}

In this section, we present the expected sensitivity to the invisible
spin-0 and spin-1 mediator hypotheses. We first determine the reach
obtained from the inclusive event yield and study its dependence on
the assumed uncertainty on the overall background normalization. We
then assess the additional sensitivity provided by differential
information using the $\mttvis$ and $D_{\rm evt}$ distributions,
with $D_{\rm evt}$ defined in Eq.~\eqref{eq:D-event}, and investigate
the impact of different binning choices and shape uncertainties.
Finally, we compare these rate- and shape-based analyses with the
fiducial spin-correlation observable $\mathcal D(\mttvis)$ introduced
in Sec.~\ref{subsec:analysis-observables}.

\subsection{Inclusive sensitivity}
\label{subsec:inclusive-reach}

We begin with the sensitivity obtained from the total number of events
passing the complete selection. This inclusive counting experiment
provides a baseline against which the gain from differential kinematic
and spin information can be assessed.

A characteristic limitation of the inclusive analysis is the
degeneracy between the signal yield and the overall background
normalization. Neglecting finite-Monte-Carlo uncertainties for
illustration, the expected yield is
\begin{equation}
    \lambda
    =
    \left(1+f_{\rm norm}\theta_{\rm norm}\right)B
    +
    \mu S ,
\end{equation}
where $B$ and $S$ denote the total background and benchmark signal
yields after the complete selection. For the background-only Asimov
dataset, $n_{\rm A}=B$, a signal contribution can be compensated by a
downward shift of the background normalization,
\begin{equation}
    \theta_{\rm cancel}
    =
    -\frac{\mu S}{f_{\rm norm}B}.
    \label{eq:theta-cancel}
\end{equation}
The sensitivity of the inclusive analysis is therefore strongly
controlled by the ratio $\mu S/(f_{\rm norm}B)$.

In the regime where the normalization uncertainty dominates over the
statistical uncertainties, this implies approximately
\begin{equation}
    q_{\mu,\rm A}
    \simeq
    \left(
        \frac{\mu S}{f_{\rm norm}B}
    \right)^2,
\end{equation}
and hence
\begin{equation}
    \mu_{\rm lim}
    \propto
    f_{\rm norm}\frac{B}{S}.
    \label{eq:mu-limit-scaling}
\end{equation}
This scaling illustrates why an inclusive counting experiment rapidly
loses sensitivity as the uncertainty on the overall background
normalization increases.

The precise background-normalization uncertainty depends on both
experimental and theoretical effects and cannot be determined
reliably within the present phenomenological setup. We therefore
adopt $f_{\rm norm}=20\%$ as a representative reference value,
following Ref.~\cite{Haisch:2016gry}, and explicitly study the
dependence on this assumption by scanning over $f_{\rm norm}$.

\begin{figure}[t]
    \centering
    \includegraphics[width=0.8\linewidth]
    {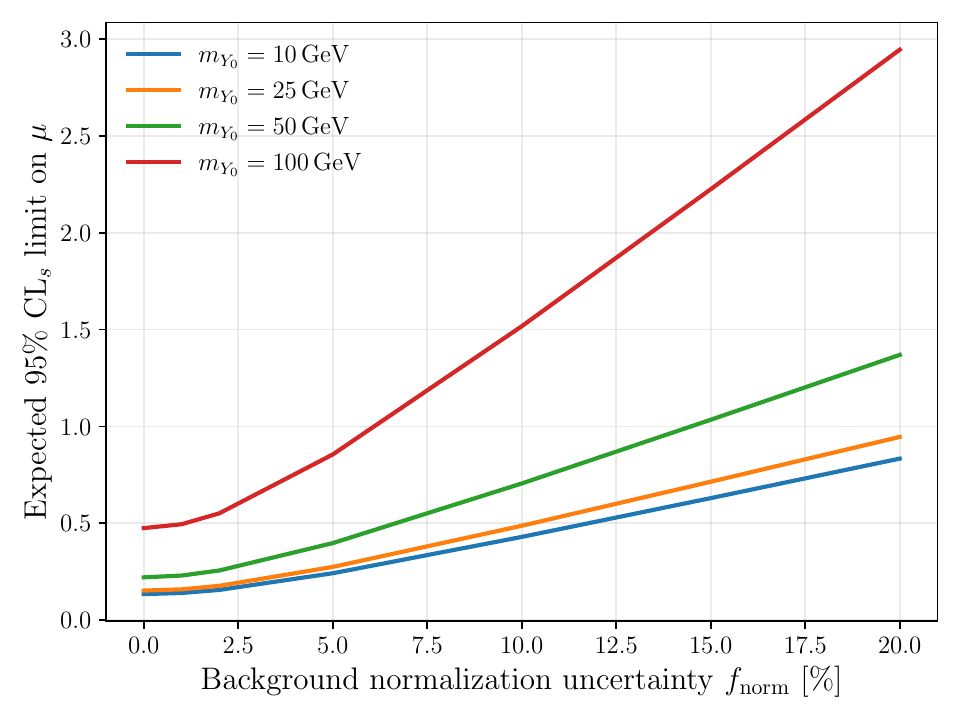}
    \caption{Median expected $95\%$ CL upper limit on the signal
    strength $\mu$ from the inclusive counting analysis as a function
    of the fractional background-normalization uncertainty
    $f_{\rm norm}$. Results are shown for the scalar-mediator
    benchmarks $m_{Y_0}=10,\,25,\,50,$ and $100~\mathrm{GeV}$. Finite-Monte-Carlo statistical uncertainties are included
    and profiled.}
    \label{fig:inclusive-reach}
\end{figure}

Figure~\ref{fig:inclusive-reach} shows that the expected limit
deteriorates as the background-normalization uncertainty increases.
At small $f_{\rm norm}$, finite data and Monte Carlo statistics become
relevant, producing the departure from the systematically dominated
behavior. At larger $f_{\rm norm}$, the approximately linear scaling
of $\mu_{\rm lim}$ is consistent with Eq.~\eqref{eq:mu-limit-scaling}.
The sensitivity also decreases with increasing mediator mass as the
selected signal yield becomes smaller.

\subsection{Sensitivity from differential distributions}
\label{subsec:differential-sensitivity}

We now investigate the improvement in sensitivity obtained by retaining
the differential information carried by the selected events. We
consider separately the visible invariant-mass distribution
$dN/d\mttvis$ and the event-level angular distribution
$dN/dD_{\rm evt}$, with $D_{\rm evt}$ defined in
Eq.~\eqref{eq:D-event}, and compare their expected limits with the
inclusive counting experiment discussed in
Sec.~\ref{subsec:inclusive-reach}.

\begin{figure}[t]
    \centering
    \includegraphics[width=\linewidth]
    {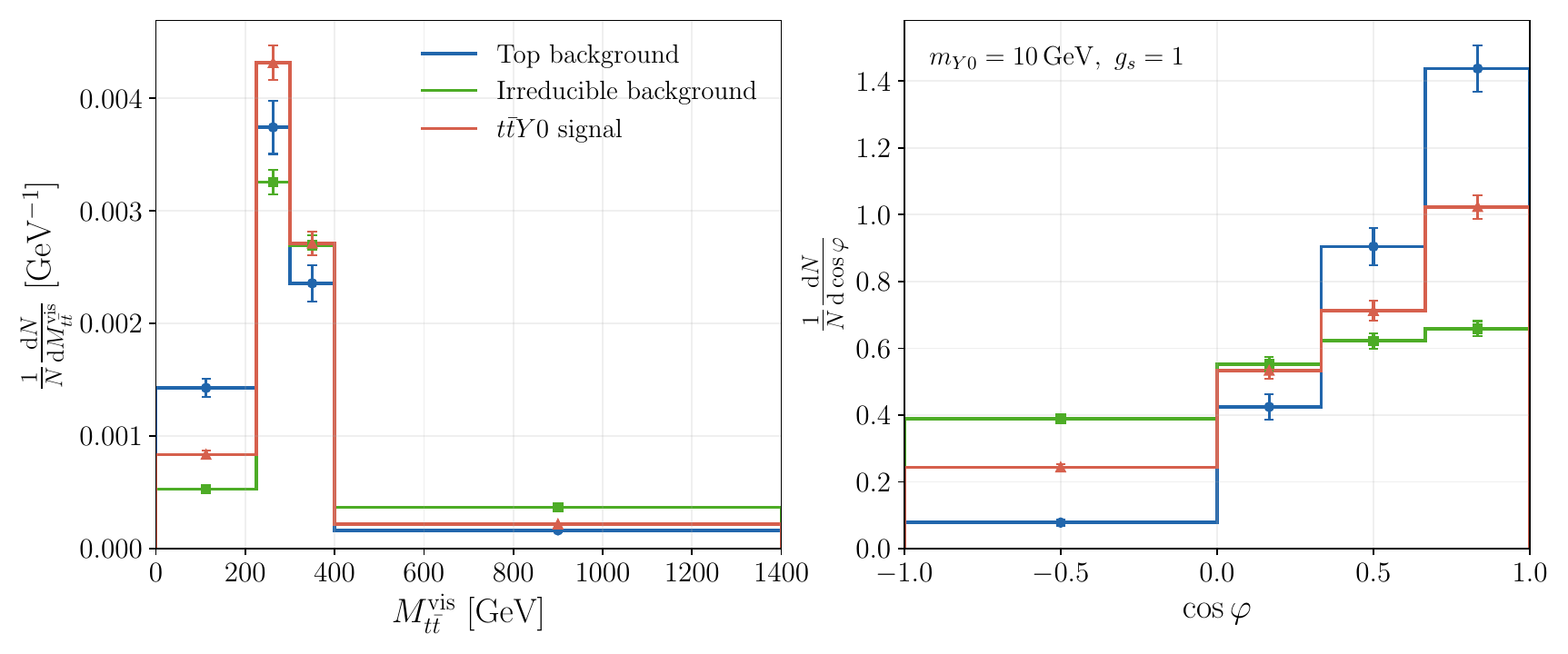}
    \caption{Normalized visible invariant-mass and angular
    distributions for the top background, the associated $t\bar tV$
    background, and the scalar-mediator benchmark
    $m_{Y_0}=10~\mathrm{GeV}$, $g_s=1$. Each process is normalized
    independently to unit area. The vertical error bars show the
    finite-Monte-Carlo statistical uncertainties for illustration; the
    statistical analysis uses the unnormalized event yields and includes
    the corresponding Monte Carlo uncertainties as described in
    Sec.~\ref{subsec:stat-methodology}.}
    \label{fig:benchmark-differential-shapes}
\end{figure}

Before presenting the statistical results, we compare the shapes of
the two distributions for the dominant top background, the associated
$t\bar tV$ background, and the scalar-mediator benchmark
$m_{Y_0}=10~\mathrm{GeV}$ with $g_s=1$. The distributions shown in
Fig.~\ref{fig:benchmark-differential-shapes} are normalized
independently to unit area in order to isolate shape differences from
the overall event rates. For the angular observable, we display the
distribution in $\cos\varphi$, which is linearly related to
$D_{\rm evt}=-3\cos\varphi$.

Both observables exhibit differences between the signal and background
templates. The visible invariant-mass distribution separates the samples mainly through their different populations in the intermediate mass regions, while the angular distribution exhibits differences between the dominant top background and the signal across the full $\cos\varphi$ range.

This motivates
testing whether the angular information provides additional sensitivity
beyond that contained in the kinematic distribution.

Both distributions are divided into four bins. For the visible
invariant mass, we use
\begin{equation}
    \mttvis:
    \qquad
    [0,225],\ [225,300],\ [300,400],\ [400,1400]~\mathrm{GeV},
    \label{eq:mttvis-binning}
\end{equation}
while for the event-level angular observable we use
\begin{equation}
    D_{\rm evt}:
    \qquad
    [-3,-2],\ [-2,-1],\ [-1,0],\ [0,3].
    \label{eq:Devt-binning}
\end{equation}
The latter corresponds to
$\cos\varphi\in[-1,0]$, $[0,1/3]$, $[1/3,2/3]$, and $[2/3,1]$. The non-uniform binning is chosen to retain sensitivity
to the signal-background shape differences while avoiding bins with
very small effective Monte Carlo populations.

For the differential analyses, the background-normalization uncertainty
is fixed to the reference value $f_{\rm norm}=20\%$. To assess the
robustness of the additional shape information, we vary the shape
uncertainty introduced in Sec.~\ref{subsec:stat-methodology} over
\begin{equation}
    f_{\rm shape}=0\text{--}50\%.
\end{equation}
This scan is intended to quantify the dependence of the sensitivity on
the assumed knowledge of the background shape rather than to model a
specific experimental systematic uncertainty. In particular,
reconstruction-induced migrations in the angular distribution would
ultimately require a dedicated detector-level treatment.

\begin{figure}[t]
    \centering
    \includegraphics[width=\linewidth]
    {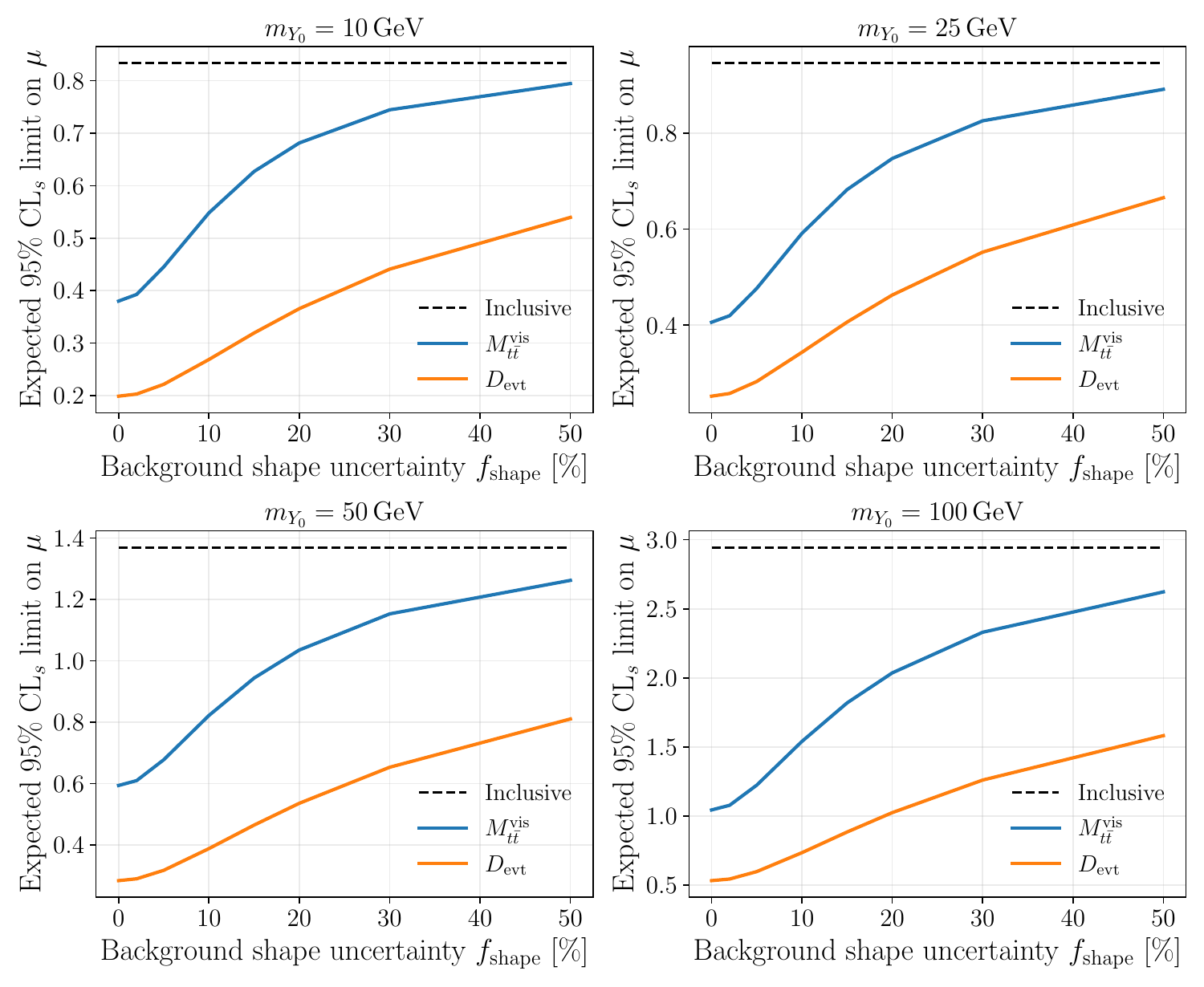}
    \caption{Median expected $95\%$ CL upper limit on the signal
    strength $\mu$ as a function of the background shape uncertainty
    $f_{\rm shape}$. Results are shown for the inclusive counting
    experiment and for the four-bin $\mttvis$ and $D_{\rm evt}$
    distributions, for $m_{Y_0}=10,\,25,\,50,$ and $100~\mathrm{GeV}$. The background-normalization uncertainty is fixed to
    $f_{\rm norm}=20\%$, and finite-Monte-Carlo statistical
    uncertainties are included.}
    \label{fig:differential-shape-reach}
\end{figure}

The results are shown in Fig.~\ref{fig:differential-shape-reach}.
Since the shape uncertainty only redistributes background events among
the bins, the inclusive limit is independent of $f_{\rm shape}$. The
limits obtained from the differential distributions instead become
progressively weaker as $f_{\rm shape}$ increases, as the background
templates acquire greater freedom to reproduce the signal-induced
shape differences.

For all mediator masses considered, the $D_{\rm evt}$ distribution
provides a stronger constraint than $dN/d\mttvis$ when the same shape
uncertainty is assigned to the two observables. This indicates that the
angular distribution contains discriminating information beyond the
overall signal enhancement and the visible invariant-mass spectrum.

The advantage of the angular observable persists under more
conservative assumptions on its shape uncertainty. Across the
benchmarks considered, the limit obtained from $D_{\rm evt}$ with
$f_{\rm shape}=50\%$ remains stronger than that obtained from
$\mttvis$ with $f_{\rm shape}=20\%$. Within the present truth-assisted
analysis, the angular information therefore remains competitive even
when substantially larger shape variations are allowed. A quantitative
experimental comparison would, however, require process-dependent
systematic variations and a realistic treatment of reconstruction
effects.

\subsection{Sensitivity from fiducial spin correlations}
\label{subsec:fiducial-spin-sensitivity}

The results of the previous subsection show that the
$dN/dD_{\rm evt}$ distribution provides stronger expected constraints
than $dN/d\mttvis$, indicating that the angular correlations contain
discriminating information beyond that encoded in the visible
invariant-mass spectrum. We therefore investigate the sensitivity of
the fiducial spin-correlation observable
$\mathcal D(\mttvis)$ introduced in
Sec.~\ref{subsec:analysis-observables}.

\begin{figure}[t]
\centering
\includegraphics[width=0.80\linewidth]
{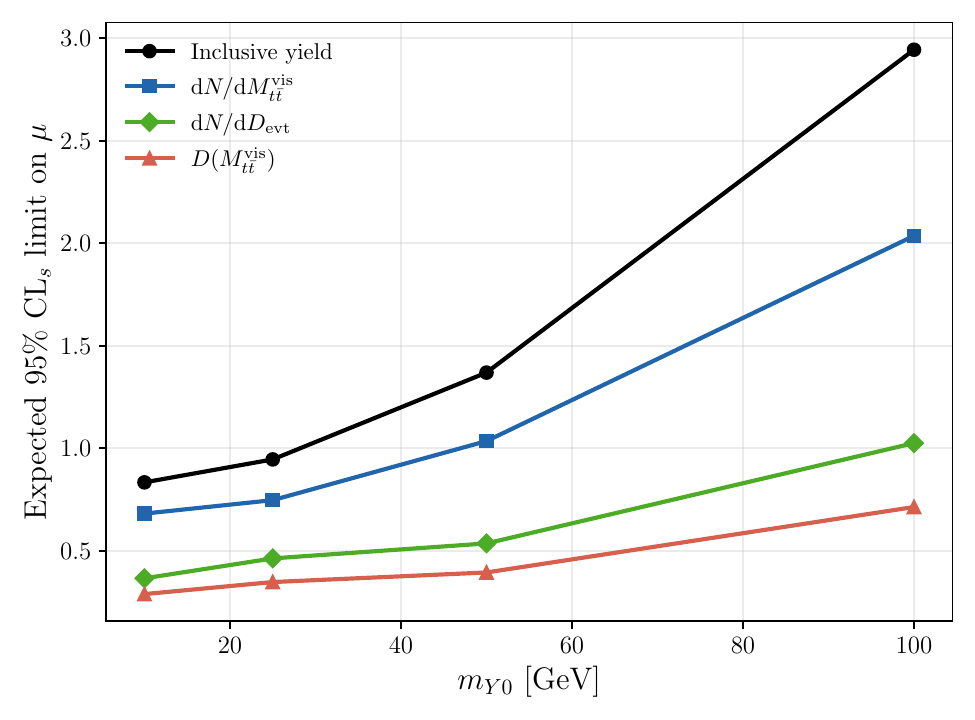}
\caption{
Expected $95\%~\mathrm{CL}_s$ upper limits on the signal strength
as a function of the scalar-mediator mass. Results are shown for the
inclusive yield, the $dN/d\mttvis$ and $dN/dD_{\rm evt}$
distributions, and the fiducial spin correlation
$\mathcal D(\mttvis)$. The background-normalization uncertainty is
fixed to $20\%$. A shape-only uncertainty of $20\%$ is assigned to
the differential event-yield distributions, while an absolute
systematic uncertainty $\delta\mathcal D_{\rm syst}=0.015$ is assigned
independently to each $\mttvis$ bin of the fiducial spin-correlation
analysis. Monte Carlo statistical uncertainties are included.
}
\label{fig:analysis-comparison}
\end{figure}

The comparison is performed using the Gaussian likelihood described in
Sec.~\ref{subsec:stat-methodology}. We fix the background-normalization
uncertainty to $f_{\rm norm}=20\%$ and, following
Ref.~\cite{Maltoni2024Quantum}, adopt an absolute systematic uncertainty
$\delta\mathcal D_{\rm syst}=0.015$, treated as uncorrelated between
the $\mttvis$ bins. For the $dN/d\mttvis$ and $dN/dD_{\rm evt}$
analyses, the shape uncertainty is fixed to $f_{\rm shape}=20\%$. The results are shown in Fig.~\ref{fig:analysis-comparison}.

For all four mediator masses, the fiducial spin-correlation analysis
provides the strongest expected constraint among the observables
considered. This shows that the spin-dependent angular information is
not fully captured by either the total event rate or the visible
invariant-mass distribution. The improvement with respect to
$dN/dD_{\rm evt}$ further indicates that retaining the dependence of
the average spin correlation on $\mttvis$ provides additional
discriminating information.

The benchmark value $\delta\mathcal D_{\rm syst}=0.015$ is motivated
by existing studies of spin correlations in standard $t\bar t$
production. Its applicability to the present topology is, however,
not guaranteed, since the additional invisible momentum and the
truth-assisted reconstruction may lead to larger experimental and
modeling uncertainties. We therefore test the robustness of the
result by varying $\delta\mathcal D_{\rm syst}$ for the benchmark point
$m_{Y_0}=10~\mathrm{GeV}$.

The purpose of this comparison is not to establish an intrinsic
hierarchy between $\mathcal D(\mttvis)$ and $dN/dD_{\rm evt}$, whose
relative performance necessarily depends on the corresponding
systematic assumptions. Rather, our aim is to provide a phenomenological
estimate of the improvement that can be obtained by exploiting
observables carrying spin-correlation information, compared with the
more conventional yield- and kinematics-based analyses. In this sense,
both $\mathcal D(\mttvis)$ and $dN/dD_{\rm evt}$ belong to the class of
spin-inspired observables whose potential sensitivity we seek to assess.

\begin{figure}[t]
\centering
\includegraphics[width=0.75\linewidth]
{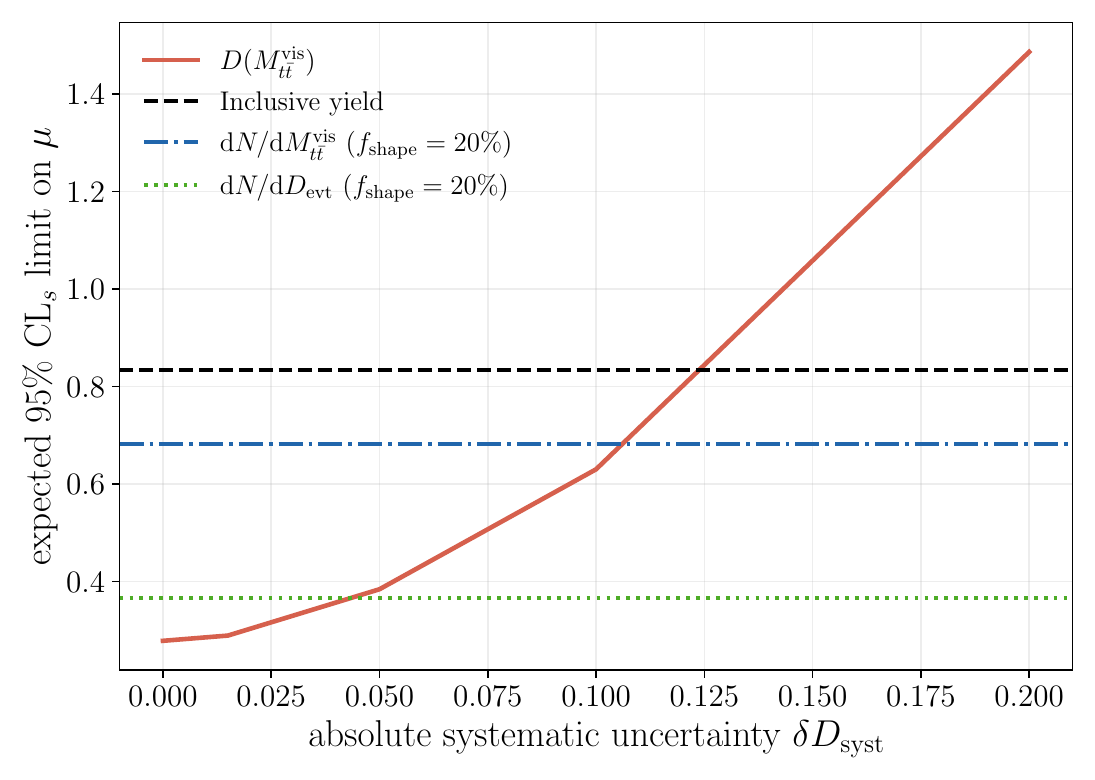}
\caption{
Expected $95\%~\mathrm{CL}_s$ upper limit on the signal strength for
the benchmark point $m_{Y_0}=10~\mathrm{GeV}$, obtained
from $\mathcal D(\mttvis)$ as a function of the assumed absolute
systematic uncertainty $\delta\mathcal D_{\rm syst}$. The horizontal
lines show the corresponding limits from the inclusive yield,
$dN/d\mttvis$, and $dN/dD_{\rm evt}$. For the differential
event-yield distributions, $f_{\rm shape}=20\%$ is assumed. Monte
Carlo statistical uncertainties and a background-normalization
uncertainty of $20\%$ are included throughout.
}
\label{fig:D-fiducial-robustness}
\end{figure}

As shown in Fig.~\ref{fig:D-fiducial-robustness}, the sensitivity of
$\mathcal D(\mttvis)$ progressively deteriorates as
$\delta\mathcal D_{\rm syst}$ increases. For moderate uncertainties,
the fiducial spin-correlation analysis remains more sensitive than the
inclusive and differential event-yield analyses. For sufficiently
large uncertainties, this advantage is gradually lost.

The gain provided by $\mathcal D(\mttvis)$ is therefore not restricted
to the nominal systematic assumption, although its quantitative impact
depends on the experimental precision with which the fiducial spin
correlations can be measured. A realistic assessment will ultimately
require a dedicated reconstruction study and a determination of the
systematic covariance between the $\mttvis$ bins.

\subsection{Spin-one mediator}
\label{subsec:spin-one-results}

We finally consider the spin-one mediator $Y_1$, focusing on the purely
vector interaction, $g_V\neq0$ and $g_A=0$. The event selection,
observable definitions, binning, and statistical treatment are kept
unchanged with respect to the spin-zero analysis. The vector benchmarks
are evaluated at the reference coupling $g_V^{\rm ref}\simeq0.70$,
defined in Eq.~\eqref{eq:vector-reference-coupling} to match the
effective top-mediator coupling of the scalar benchmark with $g_s=1$.

\begin{figure}[t]
\centering
\includegraphics[width=0.75\linewidth]
{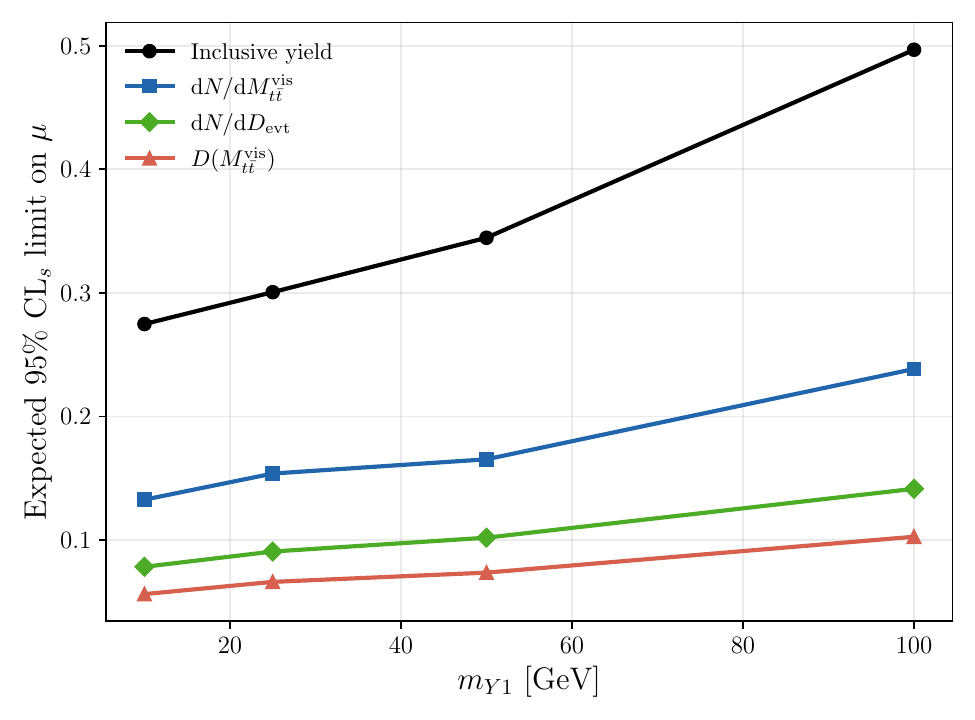}
\caption{
Expected $95\%~\mathrm{CL}_s$ upper limit on the signal strength
for a purely vector spin-one mediator as a function of $m_{Y_1}$. 
Results are shown for the inclusive yield, the $dN/d\mttvis$ and
$dN/dD_{\rm evt}$ distributions, and the fiducial spin-correlation
observable $\mathcal D(\mttvis)$.
}
\label{fig:analysis-comparison-Y1}
\end{figure}

Figure~\ref{fig:analysis-comparison-Y1} summarizes the expected
sensitivity for the four vector-mediator benchmark masses. As for the
spin-zero case, we compare the inclusive event yield, the
$dN/d\mttvis$ and $dN/dD_{\rm evt}$ distributions, and the fiducial
spin-correlation observable $\mathcal D(\mttvis)$. The reference
systematic assumptions are kept unchanged, with
$f_{\rm norm}=20\%$, $f_{\rm shape}=20\%$ for the differential
event-yield analyses, and
$\delta\mathcal D_{\rm syst}=0.015$ for the fiducial spin-correlation
analysis.

The same qualitative hierarchy observed for the spin-zero mediator is
found across the mass range considered. The inclusive counting
analysis provides the weakest constraint, while the use of
differential information improves the sensitivity. The
$D_{\rm evt}$ distribution performs better than $\mttvis$, and the
strongest limits are obtained from $\mathcal D(\mttvis)$. This shows
that the spin-dependent angular information remains discriminating
also for the vector-mediator hypothesis.

The sensitivity decreases with increasing mediator mass as the signal
rate becomes smaller, while the relative advantage of the
spin-sensitive observables is preserved. At the matched reference
coupling, the vector signal also exhibits a larger selection efficiency
than the scalar signal, particularly after the $E_T^{\rm miss}$ and
$m_{T2}^{\ell\ell}$ requirements, as seen in
Table~\ref{tab:cutflow}. This enhances the overall sensitivity to the
spin-one mediator but does not alter the relative comparison among the
four analysis strategies.

The dependence on the background-normalization and shape uncertainties
was also found to be qualitatively similar to that of the spin-zero
case. The corresponding scans are therefore not shown.

\subsection{Probing the mediator coupling structure}
\label{subsec:coupling-structure}

The exclusion reaches presented above quantify the sensitivity to the
presence of a BSM contribution. We now investigate whether the same
observables can also discriminate between different Lorentz structures
of the top--mediator interaction. We focus on the scalar and
pseudoscalar couplings of the spin-zero mediator \(Y_0\).

For a fixed mediator mass, we consider the purely scalar and
pseudoscalar benchmark hypotheses
\begin{align}
H_S &: \qquad g_s=1,\quad g_p=0,
\\
H_P &: \qquad g_s=0,\quad g_p=1.
\end{align}
The corresponding signal templates are denoted by \(S_b\) and \(P_b\),
respectively. Since the two hypotheses are non-nested, the usual
asymptotic relation between a profile-likelihood-ratio test statistic
and a universal \(\chi^2\) distribution cannot in general be assumed.
We therefore calibrate the discrimination test directly using
pseudoexperiments. Each pseudoexperiment includes both the primary measurement and the auxiliary measurements constraining the nuisance parameters, following the frequentist treatment of nuisance parameters discussed in \cite{ParticleDataGroup:2024cfk}.

For each value of the signal strength \(\mu\), the hypotheses being
compared are
\begin{equation}
H_S(\mu):\qquad B+\mu S,
\qquad
H_P(\mu):\qquad B+\mu P,
\label{eq:cp-shared-strength-hypotheses}
\end{equation}
such that the same signal strength is imposed under the scalar and
pseudoscalar hypotheses. The test is performed separately in the two
directions, \(S\to P\) and \(P\to S\), since the two benchmarks
generally predict different production rates, selection efficiencies,
and differential distributions.

For a pseudoexperiment generated under a truth hypothesis \(H_i\), we
define the signed profile-likelihood-ratio statistic
\begin{equation}
q_{i\to j}(\mu)
=
-2\log
\frac{
\mathcal L_j
\left(
x^{\rm toy},a^{\rm toy};
\mu,
\widehat{\widehat{\boldsymbol{\theta}}}_j
\right)
}{
\mathcal L_i
\left(
x^{\rm toy},a^{\rm toy};
\mu,
\widehat{\widehat{\boldsymbol{\theta}}}_i
\right)
},
\qquad
i,j=S,P,
\qquad
i\neq j,
\label{eq:coupling-hypothesis-test}
\end{equation}
where \(x^{\rm toy}\) denotes the main pseudo-dataset and
\(a^{\rm toy}\) the auxiliary measurements entering the nuisance
constraints. The nuisance parameters are profiled independently under
the two hypotheses, while the same pseudoexperiment and the same value
of \(\mu\) are used in the numerator and denominator.

Each pseudoexperiment includes fluctuations of both the main
measurement and the auxiliary measurements constraining the nuisance
parameters. The true nuisance values are kept at their nominal values.
For the event-yield observables, the main data are generated according
to the corresponding Poisson model, while normalization, shape, and
finite-Monte-Carlo uncertainties are incorporated through constrained
nuisance parameters. For the fiducial correlation
\(\mathcal D(\mttvis)\), the main pseudo-data are generated from the
Gaussian distribution associated with the data statistical covariance
only, while the finite-Monte-Carlo uncertainties on the angular means
and the additive systematic uncertainty on \(\mathcal D\) are treated
as explicit constrained nuisance parameters.

At each \(\mu\), two independent ensembles of pseudoexperiments are
generated, one under each hypothesis. For the \(i\to j\) test, the
median test statistic under the truth hypothesis is
\begin{equation}
q_{\rm med}^{\,i}(\mu)
=
\operatorname{median}
\left[
q_{i\to j}(\mu)\mid H_i
\right],
\label{eq:cp-q-median}
\end{equation}
while the critical value under the alternative hypothesis is defined
by
\begin{equation}
P_j
\left(
q_{i\to j}\geq q_{95}^{\,j}(\mu)
\right)
=
0.05.
\label{eq:cp-q95}
\end{equation}
Equivalently, the expected tail probability of the alternative
hypothesis is
\begin{equation}
p_j(\mu)
=
P_j
\left(
q_{i\to j}\geq q_{\rm med}^{\,i}(\mu)
\right).
\label{eq:cp-pvalue}
\end{equation}
We define the median expected \(95\%\) separation reach as the smallest
signal strength satisfying
\begin{equation}
p_j(\mu)\leq0.05,
\label{eq:coupling-separation-threshold}
\end{equation}
or, equivalently,
\begin{equation}
q_{\rm med}^{\,i}(\mu)
\geq
q_{95}^{\,j}(\mu).
\end{equation}
A smaller value of the separation reach therefore corresponds to a
stronger discrimination power.

We compare the inclusive yield, \(dN/d\mttvis\),
\(dN/dD_{\rm evt}\), and the fiducial spin correlation
\(\mathcal D(\mttvis)\). The first three observables use the event yields
directly as measured quantities and can therefore exploit both differences
in the total scalar and pseudoscalar production rates and differences in
their differential distributions.

The observable \(\mathcal D(\mttvis)\), instead, uses the fiducial angular
expectation values as the measured quantities. The event yields still enter
the prediction through the relative signal--background composition in each
\(\mttvis\) bin, and therefore through their dependence on the signal strength
\(\mu\). They also determine the statistical precision with which the
correlation can be measured. Thus, the yields are not discarded, but enter
as inputs to the predicted spin correlation rather than as the observables
being directly fitted. This is also why an overall normalization uncertainty
has a weaker impact on \(\mathcal D(\mttvis)\) than on the event-yield
distributions.

The reference systematic uncertainties introduced in
Sec.~\ref{subsec:stat-methodology} are retained, with
\(f_{\rm norm}=20\%\), \(f_{\rm shape}=20\%\), and
\(\delta\mathcal D_{\rm syst}=0.015\).

\begin{figure}
\centering
\includegraphics[width=\linewidth]
{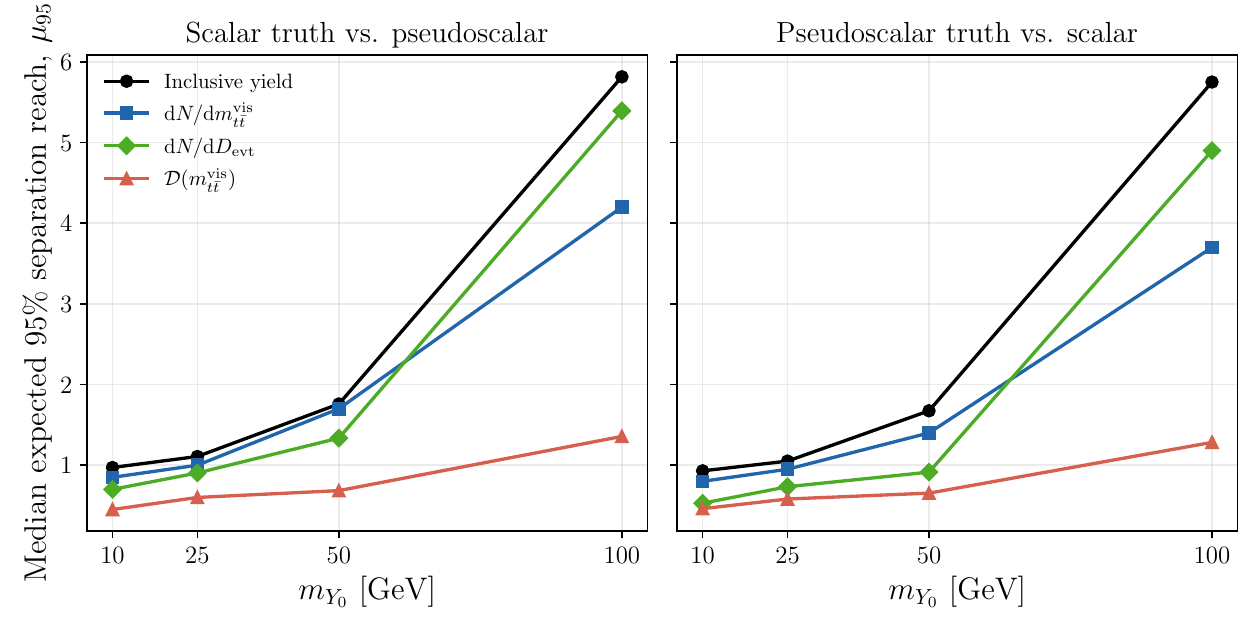}
\caption{
Median expected \(95\%\) separation reach between the scalar and
pseudoscalar coupling hypotheses as a function of the spin-zero
mediator mass. The left panel shows scalar pseudo-data tested against
the pseudoscalar hypothesis, while the right panel shows the reverse
comparison. The same signal strength \(\mu\) is imposed under the two
hypotheses, and the likelihood-ratio distributions are calibrated
using pseudoexperiments. Smaller values of the reach correspond to
stronger discrimination power.
}
\label{fig:cp-rate-shape-comparison}
\end{figure}

The results are shown in
Fig.~\ref{fig:cp-rate-shape-comparison}. The scalar and pseudoscalar
hypotheses can be distinguished using both differences in their event
rates and differences in their kinematic and spin-correlation
patterns. The two testing directions give similar, but not identical,
separation reaches, reflecting the different rates and distributions
predicted by the two signal hypotheses.

Among the observables considered,
\(\mathcal D(\mttvis)\) provides the strongest discrimination over the
full mediator-mass range. The improvement is already visible at low
mediator masses and becomes particularly pronounced for heavier
mediators. This shows that the fiducial spin-correlation pattern
contains information on the Lorentz structure of the interaction that
is not captured by the inclusive rate or by the event-level kinematic
distributions alone. In particular, while the discrimination power of
the rate-based observables deteriorates rapidly as the mediator mass
increases, the separation obtained from
\(\mathcal D(\mttvis)\) remains substantially stronger.

The differential distribution \(dN/dD_{\rm evt}\) also improves upon
the inclusive rate in parts of the considered mass range, indicating
that event-level spin information contributes to the discrimination
between the two coupling structures. However, the fiducial
correlation provides the largest gain, especially at high mediator
masses, where the scalar and pseudoscalar hypotheses become more
difficult to distinguish using the total rate alone.

We do not perform an analogous discrimination study between vector and
axial-vector couplings of the spin-one mediator. For a massive
spin-one mediator with axial couplings, the longitudinal mode couples
proportionally to \(g_A m_f/m_{Y_1}\) for a fermion of mass \(m_f\).
Perturbative unitarity therefore constrains the axial coupling through
\begin{equation}
m_f
\lesssim
\sqrt{\frac{\pi}{2}}\,
\frac{m_{Y_1}}{g_A}.
\label{eq:axial-unitarity-bound}
\end{equation}
For the top quark this bound becomes particularly restrictive for
\(m_{Y_1}\ll m_t\)~\cite{Kahlhoefer_2016,Kahlhoefer_2017}. A consistent
study of the axial-vector case would therefore require specifying the
additional dynamics that restore unitarity, and is left for future
work.
\section{Conclusions \& Outlook}
\label{sec:conclusions}

We have studied whether the spin correlations of top-quark pairs are sensitive to an invisible mediator produced in association with them, $pp\to t\bar t+Y_{0,1}$. Such a final state arises naturally in top-philic simplified dark-matter models. We work in the dileptonic channel at the HL-LHC and select events with $E_T^{\rm miss}>100~\mathrm{GeV}$ and $m_{T2}^{\ell\ell}>90~\mathrm{GeV}$. On this sample we compare four alternative analyses: the inclusive yield, the $\mttvis$ distribution, the distribution of the event-level estimator $D_{\rm evt}$, and the fiducial spin correlation $\mathcal D(\mttvis)$.

For both spin-0 and spin-1 mediators and all masses considered, observables exploiting angular and spin-correlation information provide an improvement over the inclusive and purely kinematic benchmarks. The inclusive yield gives the weakest reach, while $\mttvis$ improves on it. The angular distribution $dN/dD_{\rm evt}$ performs better than $\mttvis$ and remains more sensitive even when assigned a substantially larger shape uncertainty. Under the nominal systematic assumptions adopted in our analysis, the strongest expected limits are obtained from $\mathcal D(\mttvis)$. This advantage gradually decreases as the uncertainty on $\mathcal D$ is increased and is lost for sufficiently large systematic uncertainties; in our setup, comparable sensitivity is reached for values of $\delta\mathcal D_{\rm syst}$ of order three times the nominal benchmark, $\delta\mathcal D_{\rm syst}\sim0.05$.

The purpose of this comparison is therefore not to establish an intrinsic hierarchy between $\mathcal D(\mttvis)$ and $dN/dD_{\rm evt}$, whose relative performance depends on the corresponding systematic assumptions. Rather, it is to quantify, within a phenomenological setup, the potential improvement obtained by including observables sensitive to the angular and spin structure of the $t\bar t$ system. In this respect, both $dN/dD_{\rm evt}$ and $\mathcal D(\mttvis)$ consistently outperform the more conventional yield- and kinematics-based observables considered here.

Beyond improving the sensitivity to an invisible BSM contribution, spin-sensitive observables also help characterize the Lorentz structure of the interaction. For the spin-zero mediator $Y_0$, the fiducial spin correlation $\mathcal D(M_{t\bar t}^{\rm vis})$ provides the strongest scalar--pseudoscalar discrimination within the systematic assumptions considered, outperforming both the inclusive rate and the event-level differential observables over the mediator-mass range studied. This shows that the sensitivity is not determined by differences in the total production rate alone. Taken together, these results indicate that observables inspired by the spin structure of the $t\bar t$ system can provide complementary and potentially enhanced sensitivity to invisible new physics compared with standard rate- and kinematics-based analyses.

These conclusions are subject to the limitations of a phenomenological, truth-assisted study. The top-quark momenta are taken from the generator record, and no detector simulation is performed. Moreover, the observable used is a fiducial quantity. Once selection cuts are applied, $\mathcal D$ is an acceptance-weighted average and no longer corresponds directly to the trace of the $t\bar t$ spin-correlation matrix. A realistic assessment of the achievable sensitivity will therefore require a dedicated reconstruction study and a determination of the relevant experimental and theoretical systematic uncertainties and their correlations.

In this work $\mathcal D$ has been used purely as a discriminant between the SM and SM$+$BSM hypotheses. It is, however, directly connected to entanglement. In the inclusive limit it reduces to $D$, which measures the overlap of $\rho_{t\bar t}$ with the spin singlet, and $D<-1/3$ is a sufficient condition for entanglement. A natural next step is therefore to unfold the fiducial measurement to the inclusive production density matrix, in bins of the $t\bar t$ kinematics. It would then be possible to ask whether, and where, the presence of an invisible mediator changes the entanglement of the $t\bar t$ pair, and not only its spin correlations.

This question is of particular interest for distinguishing mediator spins. As discussed in the introduction, a spin-0 mediator modifies $\rho_{t\bar t}$ only through the kinematics of the recoiling pair and the Lorentz structure of the production amplitude. At fixed phase-space point and initial-state helicities, the reduced $t\bar t$ spin state remains pure. A spin-1 mediator, in contrast, can be entangled with the $t\bar t$ spins through its polarization. Tracing it out can then genuinely reduce the entanglement of the reduced state. Entanglement-sensitive observables could therefore help discriminate between $Y_0$ and $Y_1$, beyond the differences in rates and fiducial correlations studied here. We have not analysed this possibility in detail. Doing so would require separating this effect from the classical mixing induced by phase-space integration. It would also require accounting for traced-out spin-1 states already present in the SM, such as real gluon emission in $t\bar t$ production and the $Z$ boson in $t\bar tZ$.

$D$ is only an entanglement witness: it probes the singlet component of $\rho_{t\bar t}$ and uses only the trace of the spin-correlation matrix. The full set of coefficients $\mathcal S_{ij}$, and the polarizations if absorptive or CP-violating effects are considered, contains more information. This information can be exploited either directly in the likelihood or through quantities such as the concurrence~\cite{Wootters:1997id}, which quantifies the entanglement of an arbitrary two-qubit state. Quantities such as magic, which characterises how far a state is from being efficiently simulable on a classical computer~\cite{Veitch:2013qwv,White:2024nuc,Aoude:2025jzc} and has recently been measured in $t\bar t$ events~\cite{CMS:2025magic}, offer a further handle.

More broadly, this work illustrates how quantum information can contribute to collider phenomenology beyond testing quantum mechanics at high energies. Quantities such as entanglement markers are usually studied as properties of the $t\bar t$ state for their own sake. Here they have proven to be sensitive probes of new physics that remains out of reach of rate and kinematic measurements. They are physically motivated, can be independent of the choice of reference frame, and sensitive to the full spin structure of the production amplitude. For these reasons they provide a principled way to organize the information encoded in angular distributions, rather than an arbitrary choice of variables. At the same time, the same quantities also characterise the quantum structure of the underlying interaction. A search for invisible particles built on them would therefore deliver, as a by-product, information on how much entanglement and quantum complexity a fundamental process generates. As measurements of quantum observables at the LHC become more precise, we expect this dual role to make quantum information an increasingly valuable tool in the search for physics beyond the Standard Model.

\vfill
\acknowledgments
F.M. acknowledges A. Wulzer for useful discussions and comments during the preparation of this work.

This work is part of the doctoral thesis of F.M., within the framework of the Doctoral Program in Physics of the Autonomous University of Barcelona.
The work of F.M. is supported by the Joan Oró predoctoral grant program of the Department of Research and Universities of the Government of Catalonia and co-funded by the European Social Fund Plus (Project reference: 2024 FI-1 00715). 
A.C.-L. acknowledges the grant RYC2022-037769-I funded by MICIU/ AEI/ 10.13039/ 501100011033 and by “ESF+".
This publication is part of the R\&D\&i project PID2023-146686NB-C31 funded by MICIU/AEI/10.13039/501100011033/ and by ERDF/EU.
IFAE is partially funded by the CERCA program of the Generalitat de Catalunya.
This work is supported by ERC grant ERC-2024-SYG 101167211 funded by the European Union. Views and opinions expressed are however those of the author(s) only and do not necessarily reflect those of the European Union or the European Research Council Executive Agency. Neither the European Union nor the granting authority can be held responsible for them.
D.B. acknowledges financial support from the Spanish Ministry of Science and Innovation (MICINN) through the Spanish State Research Agency, under Severo Ochoa Centres of Excellence Programme 2025-2029 (CEX2024001442-S).
This work is an extension from the master thesis from B.B. \cite{Belmonte2026Entanglement}.

\appendix

\section{Kinematic reconstruction of the \texorpdfstring{$t\bar t$}{ttbar} system}
\label{app:reco}

In the dileptonic channel, the top and antitop quarks decay as
\begin{equation}
t\to b\,W^+(\to \ell^+\nu),
\qquad
\bar t\to \bar b\,W^-(\to \ell^-\bar\nu),
\end{equation}
so that the visible final state consists of two $b$-jets and two charged
leptons, while the two neutrinos escape detection. The definition of the
helicity basis $\{\hat k,\hat r,\hat n\}$ introduced in
Eq.~\eqref{eq:hel-basis} requires the reconstruction of the top and
antitop four-momenta and therefore of the full $t\bar t$ kinematics.

For the standard dileptonic $t\bar t$ topology, the unknown quantities
can be taken to be the three-momenta of the two neutrinos,
\begin{equation}
\vec p_{\nu}
=
(p_{\nu,x},p_{\nu,y},p_{\nu,z}),
\qquad
\vec p_{\bar\nu}
=
(p_{\bar\nu,x},p_{\bar\nu,y},p_{\bar\nu,z}),
\end{equation}
corresponding to six unknown momentum components. The neutrino energies
are subsequently fixed by the massless conditions
$E_\nu=|\vec p_\nu|$ and $E_{\bar\nu}=|\vec p_{\bar\nu}|$.

The transverse components are constrained by the measured missing
transverse momentum. If the two neutrinos are the only invisible
particles in the event,
\begin{equation}
p_{\nu,x}+p_{\bar\nu,x}
=
p_x^{\rm miss},
\qquad
p_{\nu,y}+p_{\bar\nu,y}
=
p_y^{\rm miss}.
\label{eq:standard-met-reco}
\end{equation}
Four additional constraints follow from the $W$-boson and top-quark
mass-shell conditions,
\begin{align}
(p_{\ell^+}+p_\nu)^2 &= m_W^2,
&
(p_b+p_{\ell^+}+p_\nu)^2 &= m_t^2,
\nonumber\\
(p_{\ell^-}+p_{\bar\nu})^2 &= m_W^2,
&
(p_{\bar b}+p_{\ell^-}+p_{\bar\nu})^2 &= m_t^2.
\label{eq:onshell_constraints}
\end{align}
At the idealized parton level, Eqs.~\eqref{eq:standard-met-reco} and
\eqref{eq:onshell_constraints} therefore provide six equations for the
six unknown neutrino momentum components. The resulting nonlinear
system is kinematically closed, although it may admit several discrete
solutions.

A number of strategies have been developed to reconstruct dileptonic
$t\bar t$ events. An analytic solution of the constraint equations can
be obtained by reducing the system to a quartic polynomial, as in the
Sonnenschein method~\cite{Sonnenschein:2006ud}. A geometrical
alternative is provided by the ellipse method of
Ref.~\cite{Betchart:2013nba}, in which the allowed momentum of each
neutrino is constrained to an ellipse and the measured transverse
momentum imbalance selects their intersections. Neutrino-weighting methods instead scan over otherwise unconstrained neutrino variables and assign a weight to each solution according to its compatibility with the measured missing transverse momentum~\cite{D0:1997pjc,D0:2015dxa}. The neutrino-weighting technique has also been employed by ATLAS~\cite{ATLAS:2014aus,ATLAS:2015ysm,ATLAS:2016pbv,ATLAS:2016bac}. Related weighted kinematic reconstruction methods have been used by CMS~\cite{CMS:2011acs,CMS:2012tdr}.\\
In realistic events, detector resolution, finite-width effects and combinatorial ambiguities can prevent the measured kinematics from satisfying the on-shell constraints exactly. Standard reconstruction methods therefore supplement the analytic constraints with smearing, phase-space scans or likelihood-based criteria to select among possible solutions. 

\subsection{Additional invisible particles}
\label{app:reco-extra-invisible}

The situation is qualitatively different for the processes relevant to
the present analysis. Whenever an additional invisible system $X$ is
produced together with the $t\bar t$ pair, the measured missing
transverse momentum receives contributions beyond those of the two
top-decay neutrinos,
\begin{equation}
\vec p_T^{\,\rm miss}
=
\vec p_{T,\nu}
+
\vec p_{T,\bar\nu}
+
\vec p_{T,X}.
\label{eq:extra-invisible-met}
\end{equation}
For the signal samples,
\begin{equation}
X=Y_0,Y_1,
\end{equation}
while the same issue arises for SM backgrounds containing additional
invisible particles, most notably
\begin{equation}
t\bar tZ,
\qquad
Z\to\nu\bar\nu,
\end{equation}
and for $t\bar tW$ configurations in which the additional $W$ boson
produces an unreconstructed neutrino.

If $X$ is treated as a single on-shell invisible object, its momentum
introduces three additional unknown components. The six unknown
components of the top-decay neutrinos are therefore supplemented by
$\vec p_X$, while the two missing-transverse-momentum constraints no
longer constrain the neutrino momenta alone. Equivalently, one may
write
\begin{equation}
\vec p_{T,\nu}
+
\vec p_{T,\bar\nu}
=
\vec p_T^{\,\rm miss}
-
\vec p_{T,X},
\end{equation}
but the transverse momentum of $X$ is itself unknown. The standard
dileptonic reconstruction is consequently underconstrained. If the
additional invisible system resolves into several particles, as in
$Z\to\nu\bar\nu$, the number of microscopic unknowns is even larger,
although mass-shell information associated with the parent particle
may provide additional constraints.

This observation applies not only to the irreducible SM backgrounds,
but also to the signal one aims to test. In particular, applying a
standard dileptonic $t\bar t$ solver while identifying
$\vec p_T^{\,\rm miss}$ with
$\vec p_{T,\nu}+\vec p_{T,\bar\nu}$ would assign part of the momentum of
the mediator to the top-decay neutrinos and therefore bias the
reconstructed top and antitop momenta.

In the present work, we avoid this additional reconstruction ambiguity
by taking the top- and antitop-quark four-momenta from the
generator-level hard-process record, as described in
Sec.~\ref{subsec:spin-extraction}. These momenta are used only to define
the $t\bar t$ rest frame and the helicity axes; the charged leptons
entering the angular observables are instead taken after showering and
object reconstruction. The resulting analysis is therefore
truth-assisted. Its purpose is to determine the intrinsic sensitivity
contained in the spin-correlation observables before introducing the
additional degradation associated with an experimental reconstruction
of the $t\bar t$ system.

A fully experimental implementation would require replacing this truth
information by an estimator of the top and antitop four-momenta. The
fact that Eq.~\eqref{eq:extra-invisible-met} makes the system
underconstrained does not, however, imply that no useful reconstruction
is possible. An exact event-by-event determination of all invisible
momenta is stronger than what is required for the present analysis:
the relevant question is whether the reconstructed $t\bar t$ frame
retains sufficient information about the lepton directions to preserve
the spin-correlation differences between the SM and BSM hypotheses.

\subsection{Possible reconstruction strategies}
\label{app:reco-strategies}

A reconstruction closely related to the topology studied here has
already been considered in
Ref.~\cite{Azevedo:2023xuc}. That work studies dileptonic
\begin{equation}
pp\to t\bar tY_0
\end{equation}
production with an invisible spin-zero mediator and reconstructs the
$t\bar t$ system through a kinematic fit without explicitly
reconstructing the mediator. Importantly, the authors show that, even
after including experimental effects, a meaningful reconstruction of
the $t\bar t$ system remains possible in the presence of the additional
invisible particle. This result is particularly relevant for the
present analysis, as it demonstrates that access to angular observables
of the reconstructed $t\bar t$ system need not be lost once an
additional invisible state is present.

The same reconstruction strategy has recently been extended in
Ref.~\cite{Capucha:2026oqu} to associated production with both
spin-zero and spin-one invisible mediators. This is particularly
relevant for the present work, since it closely matches the signal
topologies considered here. The authors show that the $t\bar t$ system
can still be reconstructed through a kinematic fit without explicitly
reconstructing the additional invisible mediator, and that the resulting
reconstructed kinematics retain sufficient information to study angular
observables. This provides a concrete reconstruction strategy for the problem discussed above: although the presence of the additional
invisible particle makes the system formally underconstrained, an
experimentally useful reconstruction of the $t\bar t$ system remains
possible for both spin-zero and spin-one mediator hypotheses.

A complementary possibility is to replace the exact invisible
momenta by kinematically motivated estimators. The
$M_{T2}$-Assisted On-Shell (MAOS) construction
of Ref.~\cite{Cho:2008tj}, and its application to dileptonic
$W$ decays in Ref.~\cite{Choi:2009hn}, provide an example of this
approach. In MAOS reconstruction, the transverse momenta assigned to
the invisible particles are those associated with the $M_{T2}$
minimization, while longitudinal components are inferred using
on-shell constraints. The resulting momenta need not coincide with
the true invisible momenta event by event, but can remain sufficiently
correlated with them to reconstruct parent-particle properties and
spin-sensitive observables.

The original MAOS construction assumes the usual topology with two
invisible particles and therefore does not directly solve
Eq.~\eqref{eq:extra-invisible-met}. Nevertheless, it illustrates a
useful general principle for the present problem: an underconstrained
event can be supplemented by physically motivated prescriptions that
define estimators for the unmeasured momenta. An extension to the
$t\bar t+X$ topology could, for example, treat the momentum attributed
to the additional invisible system as a nuisance quantity and select
the configuration that optimizes a kinematic or likelihood criterion.

More generally, the reconstruction may be formulated as an inference
problem rather than as the exact inversion of the event kinematics.
Given the observed objects,
\begin{equation}
\mathcal O
=
\left\{
p_{\ell^+},p_{\ell^-},
p_b,p_{\bar b},
\vec p_T^{\,\rm miss}
\right\},
\end{equation}
one may seek an estimator or probability distribution for the
unobserved top momenta,
\begin{equation}
P\left(
p_t,p_{\bar t}
\,\middle|\,
\mathcal O
\right).
\end{equation}
Constrained kinematic fits and likelihood-based methods provide one
implementation of this idea. Matrix-element methods can additionally
weight different invisible configurations using the corresponding
hard-scattering probability, while multivariate regression or modern
machine-learning methods could learn the mapping between the measured
event and the top-quark kinematics from simulated samples. In such
approaches, the extra invisible momentum need not be artificially
assigned to the two top-decay neutrinos, but can instead be marginalized
or profiled over as an additional latent degree of freedom.

For the purpose of the quantum-tomography analysis, the relevant
performance metric would not necessarily be the event-by-event
resolution of every component of $p_t$ and $p_{\bar t}$. Instead, one
should quantify how the reconstruction modifies the angular moments
from which the spin-correlation coefficients are extracted. Denoting
schematically by $R$ a given reconstruction procedure, one would
compare
\begin{equation}
\mathcal S_{ij}^{\rm reco}
=
\frac{9}{\alpha_a\alpha_b}
\left\langle
\cos\theta_{ai}^{\,R}
\cos\theta_{bj}^{\,R}
\right\rangle_{\mathcal F}
\end{equation}
with the corresponding truth-assisted quantity. The reconstruction can
induce both a bias in the central value and migrations between
$m_{t\bar t}^{\rm vis}$ bins, as well as correlations among the
different $\mathcal S_{ij}$ coefficients. These effects would have to
be calibrated on simulated samples and propagated as additional
experimental systematic uncertainties to $\mathcal S_{ij}$,
$\mathcal D$, and any observable constructed from them.

In particular, a reconstruction method need not reproduce the true top
momenta exactly in order to remain useful. What matters for the present
search is whether the difference
\begin{equation}
\Delta\mathcal S_{ij}
=
\mathcal S_{ij}^{\rm SM+BSM}
-
\mathcal S_{ij}^{\rm SM}
\end{equation}
survives the reconstruction with sufficient statistical and systematic
precision. A dedicated comparison of truth-assisted and reconstructed
spin observables for $t\bar t+$invisible events would therefore provide
the natural next step towards an experimental implementation of the
analysis.

Such a study is beyond the scope of the present work. The exclusion
reaches reported here should consequently be interpreted as estimates
of the sensitivity available from the underlying spin information
under truth-assisted reconstruction. Establishing the experimentally
achievable sensitivity will require applying a dedicated reconstruction
strategy consistently to the signal and to SM backgrounds containing
additional invisible particles, and propagating the corresponding
reconstruction uncertainties to the spin-correlation measurement.\\
The existing kinematic-fit studies of Refs.~\cite{Azevedo:2023xuc,Capucha:2026oqu} indicate, however, that the presence of the additional invisible state does not preclude a viable reconstruction of the \(t\bar t\) system.
\bibliographystyle{JHEP}
\bibliography{biblio}

\end{document}